\documentclass[submission,Phys]{SciPost}

\usepackage[utf8]{inputenc} 
\usepackage[T1]{fontenc} 	
\usepackage[english]{babel} 

\usepackage[bitstream-charter]{mathdesign}
\usepackage{geometry} 		
\usepackage{amsmath} 		
\usepackage{amsthm} 		
\usepackage{mathtools} 		
\usepackage{float} 			
\usepackage{graphicx} 		
\usepackage{tabularx} 		
\usepackage{booktabs} 		
\usepackage{color, xcolor} 	
\usepackage{pdfpages} 		
\usepackage{multirow} 		
\usepackage{multicol} 		
\usepackage{caption} 		
\usepackage{subcaption} 	
\usepackage{enumitem} 		
\usepackage{setspace} 		
\usepackage{xspace} 		
\usepackage{ragged2e} 		
\usepackage{stackrel} 		
\usepackage{tikz} 			
\usepackage{braket} 		
\usepackage{bm} 			
\usepackage{slashed} 		
\usepackage{lastpage} 		
\usepackage{cite} 			
\usepackage[normalem]{ulem} 
\usepackage{fontawesome} 	
\usepackage{tocloft} 		
\usepackage{titlesec} 		
\usepackage{doi} 			
\usepackage{hyperref} 		
\usepackage[most]{tcolorbox} 					
\usepackage[nameinlink, capitalize]{cleveref} 	
\usepackage[nottoc, notlot, notlof]{tocbibind} 	

\hypersetup{
	pdftitle={What's Anomalous in LHC Jets?},
	pdfauthor={Buss et al.},
	colorlinks=true, 			
	linkcolor={red!50!black}, 	
	citecolor={blue!50!black}, 	
	urlcolor={blue!80!black} 	
} 

\DeclareSymbolFont{usualmathcal}{OMS}{cmsy}{m}{n}
\DeclareSymbolFontAlphabet{\mathcal}{usualmathcal}

\setitemize{itemsep=0pt, parsep=0pt} 				
\setenumerate{itemsep=0pt, parsep=0pt} 				
\setitemize{itemsep=2pt,topsep=2pt,parsep=0pt,partopsep=0pt,leftmargin=*}
\setenumerate{itemsep=0pt,topsep=2pt,parsep=0pt,partopsep=0pt,labelindent=3pt,leftmargin=*}
\newcommand{\qqquad}{\qquad\quad}
\newcommand{\qqqquad}{\qquad\qquad}

\newcommand\one{\leavevmode\hbox{\small1\normalsize\kern-.33em1}}
\newcommand{\abs}[1]{\lvert#1\rvert} 		

\newcommand{\loss}{\mathcal{L}} 	

\newcommand{\arXiv}[2][]{%
	\ifthenelse{\equal{#1}{}}%
	{\href{http://arxiv.org/abs/#2}{arXiv:#2}}%
	{\href{http://arxiv.org/abs/#2}{arXiv:#2~[#1]}}}

\newcommand{\mev}{\text{MeV}}
\newcommand{\gev}{\text{GeV}}
\newcommand{\tev}{\text{TeV}}

\def\slashchar#1{\setbox0=\hbox{$#1$}           
   \dimen0=\wd0                                 
   \setbox1=\hbox{/} \dimen1=\wd1               
   \ifdim\dimen0>\dimen1                        
      \rlap{\hbox to \dimen0{\hfil/\hfil}}      
      #1                                        
   \else                                        
      \rlap{\hbox to \dimen1{\hfil$#1$\hfil}}   
      /                                         
   \fi}

\graphicspath{{./figs/}}       

\definecolor{nicered}{rgb}{0.7,0.1,0.1}
\definecolor{nicegreen}{rgb}{0.1,0.5,0.1}
\definecolor{violet}{rgb}{0.7,0.3,0.3}
\hypersetup{colorlinks,citecolor= nicegreen,linkcolor= nicered}

\begin{document}

\begin{center}{\Large \textbf{
      What's Anomalous in LHC Jets?
}}\end{center}

\begin{center}
Thorsten Buss\textsuperscript{1},   
Barry M. Dillon\textsuperscript{1}, 
Thorben Finke\textsuperscript{2}, 
Michael Kr\"amer\textsuperscript{2}, \\
Alessandro Morandini\textsuperscript{2},
Alexander M\"uck\textsuperscript{2}, 
Ivan Oleksiyuk\textsuperscript{2}, and
Tilman Plehn\textsuperscript{1}
\end{center}

\begin{center}
{\bf 1} Institut f\"ur Theoretische Physik, Universit\"at Heidelberg, Germany\\
{\bf 2} Institute for Theoretical Particle Physics and Cosmology (TTK), RWTH Aachen University, Germany \\
\end{center}

\begin{center}
\today
\end{center}

\section*{Abstract}
         {\bf Searches for anomalies are a significant motivation for the
           LHC and help define key analysis steps, including triggers. We
           discuss specific examples how LHC anomalies can be defined through
           probability density estimates, evaluated in a physics space
           or in an appropriate neural network latent space, and discuss the model-dependence
           in choosing an appropriate data parameterisation. 
           We
           illustrate this for classical k-means clustering, a
           Dirichlet variational autoencoder, and invertible neural
           networks. For two especially challenging scenarios of jets
           from a dark sector we evaluate the strengths and
           limitations of each method.}

\vspace{1pt}
\noindent\rule{\textwidth}{1pt}
\tableofcontents\thispagestyle{fancy}
\noindent\rule{\textwidth}{1pt}
\vspace{1pt}

\newpage
\section{Introduction}
\label{sec:intro}

Searches for new physics at the LHC are traditionally defined by
testing theory hypotheses and comparing them to Standard Model
predictions using likelihood methods. Weaknesses of this approach are
that search results are hard to generalize, and that we still live in
fear of having missed a discovery by not asking the right
questions. Modern machine learning (ML) offers a conceptual way out of this
dilemma through anomaly searches. Looking at LHC physics these
concepts can be developed most easily for QCD jets, analysis objects
available in huge numbers with much less physics complexity than full
events.
Nevertheless, they are complex enough such that possible new physics signatures can hide in a non-trivial way.
\medskip

Considering the task of finding anomalies in these jets, unsupervised ML methods are favored.
In contrast to supervised methods, unsupervised methods do not rely on labeled data.
This allows for sensitivity to a broad range of potential signal models and the application directly on data.
Autoencoders(AEs) are a simple unsupervised ML-tool relying on a bottleneck in a mapping of a data representation onto itself, and constructing a typical object.
They have been shown to identify anomalous jets in a
QCD jet sample, indicating that anomaly searches at the LHC are a
promising new analysis direction for the upcoming LHC
runs~\cite{Heimel:2018mkt,Farina:2018fyg}.  Autoencoders do not induce
a structure in latent space, so we have to rely on the reconstruction
error as the anomaly score. This corresponds to a definition of
anomalies as an unspecific kind of outliers. The conceptual weakness
of autoencoders becomes obvious when we invert the searches, for
instance searching for anomalous QCD jets in a top-jet
sample~\cite{Finke:2021sdf}. This failure in symmetric searches leads
us to the more fundamental question how we define anomalous jets or
events at the LHC and what kind of anomaly measure captures
them~\cite{Dillon:2021nxw}.

Moving beyond purely reconstruction-based autoencoders, variational
autoencoders~\cite{kingma2014autoencoding} (VAEs) add structure to the
latent bottleneck space. In the encoding step, a high-dimensional data
representation is typically mapped to a low-dimensional latent distribution,
from which the decoder learns to generate new high-dimensional
objects. The latent bottleneck space then contains structured
information which might not be apparent in the high-dimensional input
representation. Again, VAEs have been shown to work for anomaly
searches with LHC
jets~\cite{Cerri:2018anq,Cheng:2020dal}, and we can
replace the reconstruction loss by an anomaly score defined in latent
space. At this point, an interesting and relevant question becomes the
choice of latent space and specific anomaly score, for instance in a
Dirichlet latent
space~\cite{srivastava2017autoencoding,joo2019dirichlet} which
encourages mode separation~\cite{Dillon:2021nxw}.

ML-methods for anomaly detection
at the LHC have generally received a lot of attention in the context
of anomalous jets~\cite{ Aguilar-Saavedra:2017rzt, Roy:2019jae,
  Dillon:2019cqt,
  Aguilar-Saavedra:2020uhm,Atkinson:2021nlt,Kahn:2021drv,Dolan:2021pml,Canelli:2021aps},
anomalous events pointing to physics beyond the Standard Model~\cite{DAgnolo:2018cun, Hajer:2018kqm, Komiske:2019fks,
  Blance:2019ibf, Romao:2019dvs, DAgnolo:2019vbw, Andreassen:2020nkr,
  Amram:2020ykb, Matchev:2020wwx, Komiske:2020qhg, Romao:2020ojy,
  Dillon:2020quc, Romao:2020ocr, Khosa:2020qrz, Mikuni:2020qds,
  vanBeekveld:2020txa,Jawahar:2021vyu,Mikuni:2021nwn}, or enhancing
established search strategies~\cite{Collins:2018epr, DeSimone:2018efk,
  Collins:2019jip, Mullin:2019mmh, 1815227,
  Park:2020pak,Hallin:2021wme}.  They include a first ATLAS
analysis~\cite{Aad:2020cws}, experimental validation of some of the
methods~\cite{Komiske:2019jim, Knapp:2020dde}, quantum machine
learning~\cite{Ngairangbam:2021yma}, applications to heavy-ion
collisions~\cite{Thaprasop:2020mzp}, the DarkMachines
challenge~\cite{Aarrestad:2021oeb}, and the LHC~Olympics~2020
community
challenge~\cite{Kasieczka:2021xcg,Bortolato:2021zic}.\medskip

In spite of this wealth of possible practical applications, the
fundamental question still needs to be studied, namely \textsl{what
  defines an anomaly search at the LHC?}
For large and stochastic
datasets, the concept of outliers is difficult to define unambiguously,
because any LHC jet
or event configuration will occur with a finite probability,
especially after we include detector
imperfections~\cite{Nachman:2020lpy,Batson:2021agz,Dorigo:2021iyy,Caron:2021wmq,Fraser:2021lxm}.
In this situation, a simple, working definition of anomalous data is
an event which lies in a low-density phase space region. Such a phase
space region can be defined based on any set of kinematic observables,
notably including a latent space variable constructed by a neural
network. While such a general definition cannot be understood directly
using quantum field theory, it can be linked to first-principles
through the corresponding simulations.
An alternative definition based on overdensities can be applied for localised signals.
These signals are typically localised in some global observable like the invariant mass of the events.
The background density then needs to be inferred e.g.\ through sideband methods.
Weakly supervised methods such as the classification without labels (CWoLa) method \cite{Metodiev:2017vrx} can be applied in these settings to learn a likelihood ratio classifier for an anomalous signal in various forms\cite{Collins:2019jip,Nachman:2020lpy,Hallin:2021wme}.

For anomaly searches at the LHC we probe and encode the phase space
probability of the known background through a set of jets or
events. This training of the anomaly-search network can use
simulations of a pure background dataset, or it can use actual data
under the assumption that a small signal contamination will be ignored
by the respective network training and architecture. Once the
background density is encoded in the network, the goal of 
the anomaly detection methods discussed in this work
is to derive an anomaly score for each data point, based on
the learned background density. This way, our anomaly detection methods are fundamentally linked to density estimation, which in turn depends strongly how we define the observables in the analysis.
They are also closely
linked to standard LHC search strategies, where we test and rule out
background hypotheses without reference to a specific signal
hypothesis. Challenges to this density-based anomaly detection using
machine learning are the high dimensionality of the feature vectors
describing LHC events, or the fact that the probability densities in
the latent and phase spaces are not invariant under
reparameterizations or reweighting of the input data and the choice of
network architectures. In this paper we will study three different
ways of defining an anomaly score for LHC jets based on density
estimation, to illustrate the challenges and advantages of different
approaches and network architectures.\medskip

As reference datasets we introduce two dark-matter-inspired jet
signatures in Sec.~\ref{sec:data}. The underlying new physics model is
Hidden Valleys, made of a strongly interacting, light dark
sector~\cite{Strassler:2006im,Morrissey:2009tf,Knapen:2021eip}. Our
physics task is similar to Ref.~\cite{Canelli:2021aps}, where the
technical focus is on a fixed signal type and set of observables.
When we produce particles from this dark sector, they can decay within
the dark sector and form a dark shower, but this dark shower will
eventually switch to SM fragmentation and turn into a semi-visible
jet~\cite{Cohen:2015toa,Cohen:2017pzm,Pierce:2017taw,Beauchesne:2017yhh,
  Bernreuther:2019pfb,Bernreuther:2020vhm} or a pure, modified QCD
jet~\cite{Heimel:2018mkt}. In both cases we can use ML-based subjet
tools to tag the signal jets, provided we know and control the signal
hypothesis. The problem is that the model parameter space is too large
to rely on standard hypothesis testing; there are also reasons to doubt that the dark sector showering modelled
in Pythia~\cite{pythia} is accurate due to differences between the strong sector in the SM and 
in the dark sector. So our strategy will be
to search for such dark jets using anomaly detection on jets.  We note
that both our signals are particularly hard to distinguish from QCD
jets, unlike fat jets arising from Standard Model decays.\medskip

Facing the task of extracting our two dark jets signals from QCD jets
in an unsupervised, data-driven setup we will discuss and benchmark
three different methods.  Our first approach is based on classic
\textit{k-means clustering} combined with density estimation, and does
not require modern deep learning. In Sec.~\ref{sec:kmeans} we
introduce a general setup suitable for stochastic datasets and show how
the anomaly score can be optimized for generic classes of anomalies.  When it comes to the
best-performing latent spaces, we found that \textsl{Dirichlet VAEs}
(DVAEs) outperform for example standard VAEs or latent Gaussian mixture
models for hadronically decaying, boosted top jets as anomaly~\cite{Dillon:2021nxw}.
In Sec.~\ref{sec:dvae} we show how our
two signal jet samples benefit from a better choice in preprocessing,
and how the jet representations as jet images and energy flow
polynomials (EFPs)~\cite{Komiske:2017aww} compare in terms of anomaly
searches. 
Our third method uses \textit{normalizing flow networks}~\cite{dinh2017density, kingma2018glow},
specifically invertible neural networks (INNs)~\cite{inn,cinn},
to bijectively map
phase space to latent space. To limit the dimensionality of this
mapping we use EFPs as the phase space representation.  INNs are the
cleanest way to directly estimate the density of the jets in phase
space (or, physics space).  They learn an invertible transformation
from the phase space of a jet to a multi-dimensional Gaussian along
with a Jacobian that ensures the density is properly accounted for in
the transformation.  This gives us both a structured Gaussian latent
space for the jets, and a method to estimate the density of the jets
in the physics space.  We want to study the use of densities as
anomaly scores.  Comparing these three fundamentally different
concepts on the two dark jets samples we find very similar performance
and a sizeable dependence on the preprocessing of the respective
datasets, specifically the reweighting of the inputs.

\section{Dark jets}
\label{sec:data}

Fat jets from boosted, hadronically decaying particles are among the
most complex objects entering LHC analyses.  Produced from a single
relativistic particle and undergoing decay, showering, and
hadronization, they contain between 20 and 100 relevant constituents
per jet, in which we need to identify the crucial decay and showering
patterns.  In the Standard Model, established fat jet signatures arise
from boosted top quarks, boosted weak bosons, and boosted Higgs
bosons. In addition, many interesting new physics scenarios can lead
to fat jets.  One class of such models are Hidden Valley
models~\cite{Morrissey:2009tf,Strassler:2006im,Knapen:2021eip}.

The variety of Hidden Valley models and their parameters is extensive,
for our purpose their leading feature is a strongly coupled $SU(3)_d$
dark sector with fermions coupling to the SM through some portal.
Jets in these models can be produced from new physics states with
couplings to the SM and dark sectors, so the showering involves
radiation into the dark sector.  The resulting jets are referred to as
semi-visible jets. However, not all jets from Hidden Valley models
result in a significant showering of stable dark hadrons, so we refer
to them more generally as dark jets.  Because LHC experiments cannot
search for the full range of possible anomalous jets using
theory-based hypothesis methods, such searches are a clear candidate
for an unsupervised machine-learning treatment, similar to strategies
proposed for SUEP signals~\cite{Barron:2021btf}.  Before we proceed
with the different anomaly-search strategies, we briefly describe the
two signal datasets we use for our analysis, including possible data representations and data preprocessing. 

\subsection{Datasets}

To cover the range of challenges and solutions in density-based
anomaly searches, we define two signal benchmarks with different
underlying physics features:
\begin{enumerate}
\item Aachen dataset
  \begin{align}
    pp \to Z^\prime \to \bar{q}_dq_d
    \qquad \text{with} \qquad m_{Z^\prime} = 2~\tev \quad \text{and} \quad m_{q_d} =500~\mev \; .
  \end{align}
  The $Z^\prime$ mediator to the dark sector is described by a
  weakly interacting $U(1)^\prime$ gauge group, and the dark quarks
  are charged under a strongly coupled dark sector and this
  $U(1)^\prime$.  Like QCD, the dark sector contains dark pions,
  $m_{\pi_d} = 4$~GeV, and dark rho mesons, $m_{\rho_d} = 5$~GeV.
  The $Z^\prime - \rho_d^0$ mixing leads to the decay of the $\rho_d^0$
  to SM-quarks.  The other dark mesons are stable and
  escape the detector.  The fraction of constituents escaping the
  detector is $r_\text{inv} = 0.75$.  This scenario is a typical
  semi-visible jet, as explored in detail in~\cite{Bernreuther:2020vhm,Finke:2021sdf}.
\item Heidelberg dataset
  \begin{align}
    pp  \to   \bar{q}_dq_d
    \qquad \text{with} \qquad m_{q_d} = 50~\gev \; .
  \end{align}
  The dark quarks are charged under $SU(3)_c$ and the dark
  $SU(3)_d$, so after pair-production the dark quarks will radiate
  into the SM and dark sectors.  Eventually, the dark states decay
  back to SM particles plus a dark boson $b_v$ with mass 5~GeV.  This
  dark boson hadronizes to scalar and pseudo-scalar dark meson states
  with masses assumed to be 10~GeV.  For our choice of model parameters, the dark mesons decay back to the
  SM, i.e.\ $r_\text{inv} \simeq
  0$ for the Heidelberg dataset~\cite{Heimel:2018mkt}.
\end{enumerate}
The Aachen dataset is simulated using Madgraph5~\cite{madgraph} for
the hard process and the Hidden Valley
model~\cite{Carloni:2010tw,Carloni:2011kk} in Pythia8.2~\cite{pythia}
for showering and hadronization.  The Heidelberg dataset is simulated
using just the Hidden Valley model in Pythia8.2.
The light QCD background jets are simulated using MadGraph5 to obtain di-jet events and Pythia8.2 for showering and hadronization.
For a fast detector simulation we rely on Delphes3~\cite{delphes}. 
A background which we do not consider here arises from detector malfunctions such as dead cells, however this is not modelled by Delphes so we are unable to implement it here.  Nevertheless this does not alter the core results of the analysis.
The jets are reconstructed using the anti-$k_T$ algorithm~\cite{anti-kt} with
$R=0.8$ in FastJet~\cite{fastjet} and fulfill
\begin{align}
  p_{T,j} = 150~...~300~\gev
  \qquad \text{and} \qquad
  |\eta_j| < 2 \; .
\label{eq:kinematics}
\end{align}
Although these parameters and cuts do not guarantee that all decay products of the Heidelberg dark quarks end up in the same jet, in many cases they will.
These two signal benchmarks allow us to probe different aspects of
dark jets, the Aachen dataset providing a typical example of
semi-visible jets and the Heidelberg dataset providing a typical example 
of decay-like or two-prong-like new physics jets.

\begin{figure}[t!]
  \includegraphics[width=0.33\textwidth]{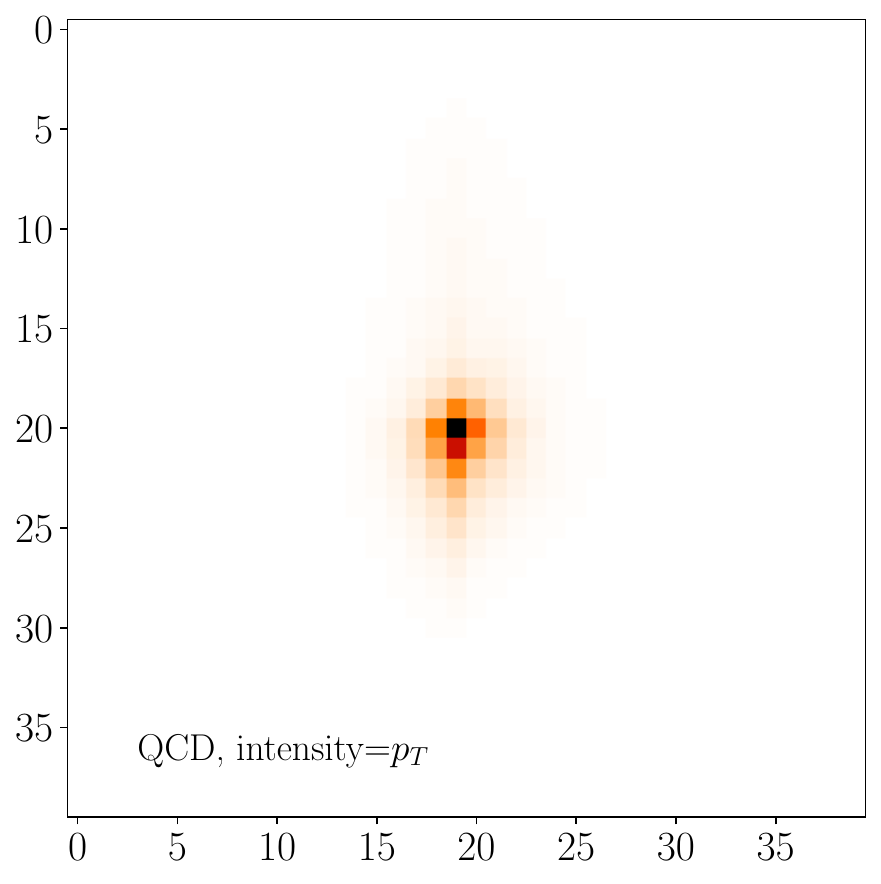}
  \includegraphics[width=0.33\textwidth]{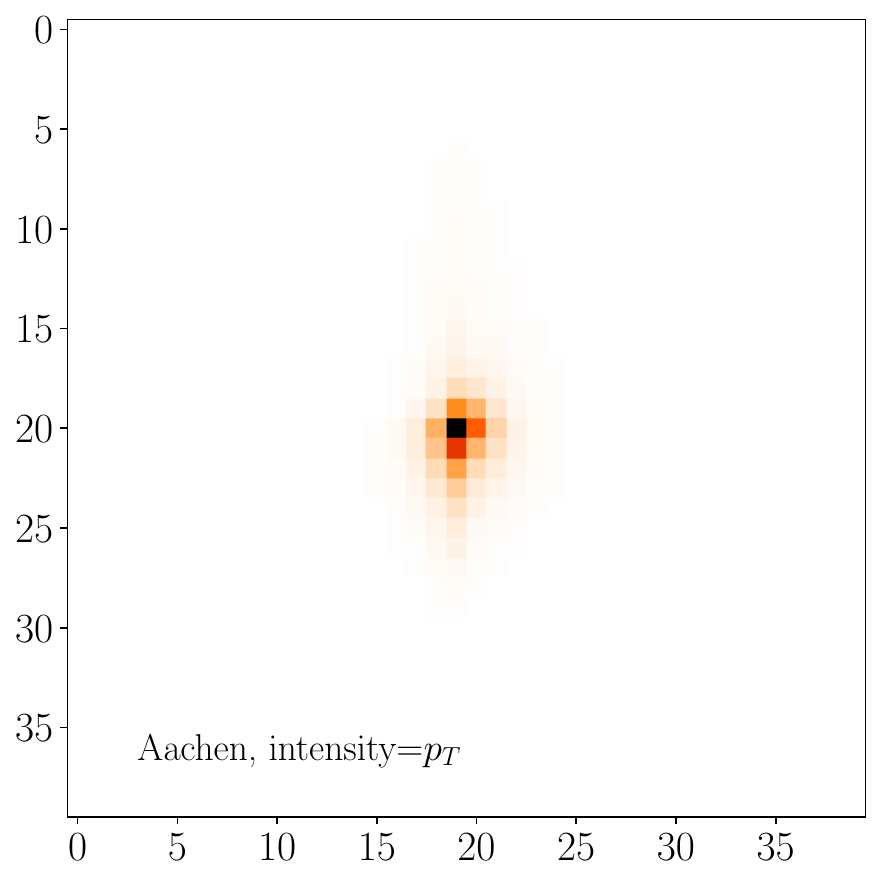}
  \includegraphics[width=0.33\textwidth]{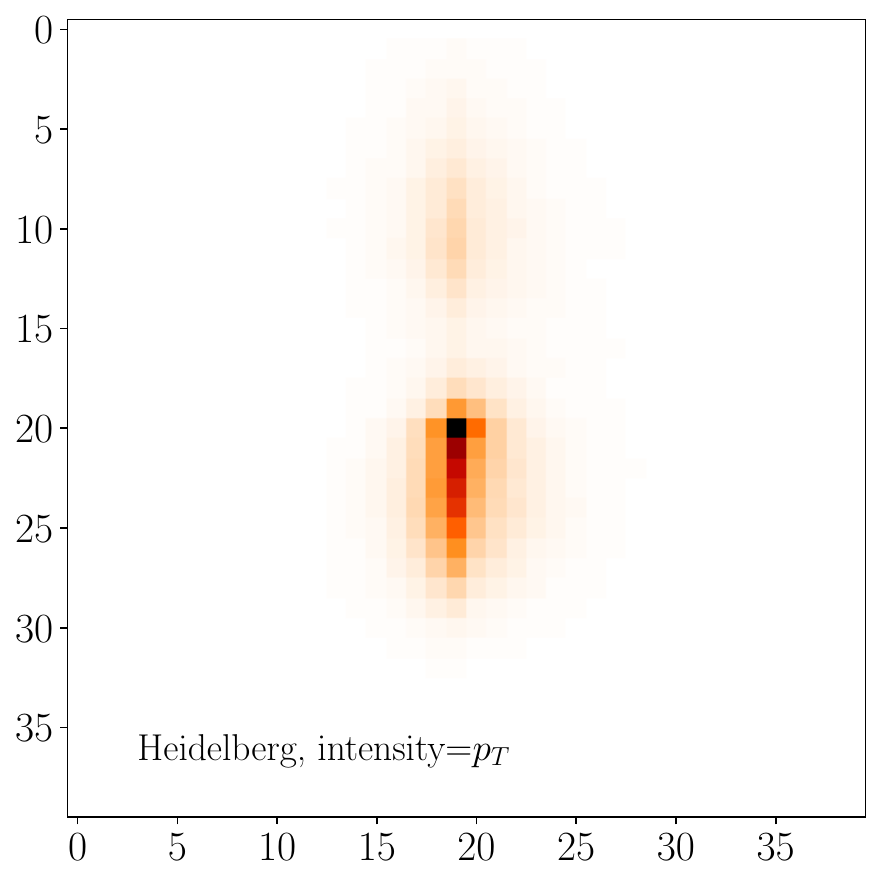}
  \caption{Averaged preprocessed jet images for the QCD background and
    for the two dark-jet samples.}
  \label{fig:dark-imgs}
\end{figure}

To represent the jets we use the standard jet images or calorimeter
images, replacing the usual high-level observables by the direct
detector output and applying basic
preprocessing~\cite{deOliveira:2015xxd,Baldi:2016fql,Komiske:2016rsd,Kasieczka:2017nvn,Macaluso:2018tck}.
The size of the images is $40 \times 40$ pixels. The key
preprocessing steps are centering and normalizing the pixels jet by
jet, to sum to one. In Fig.~\ref{fig:dark-imgs} we illustrate the different patterns
using averaged jet images after centering them and rotating the main
axis to 12 o'clock. In this representation, the semi-visible nature of
the Aachen dataset can hardly be distinguished from the QCD
background, while the Heidelberg dataset shows its two-prong structure
with a typical distance between the prongs induced by the 
window in the transverse momentum.\medskip

\begin{figure}[t]
  \includegraphics[width=0.33\textwidth]{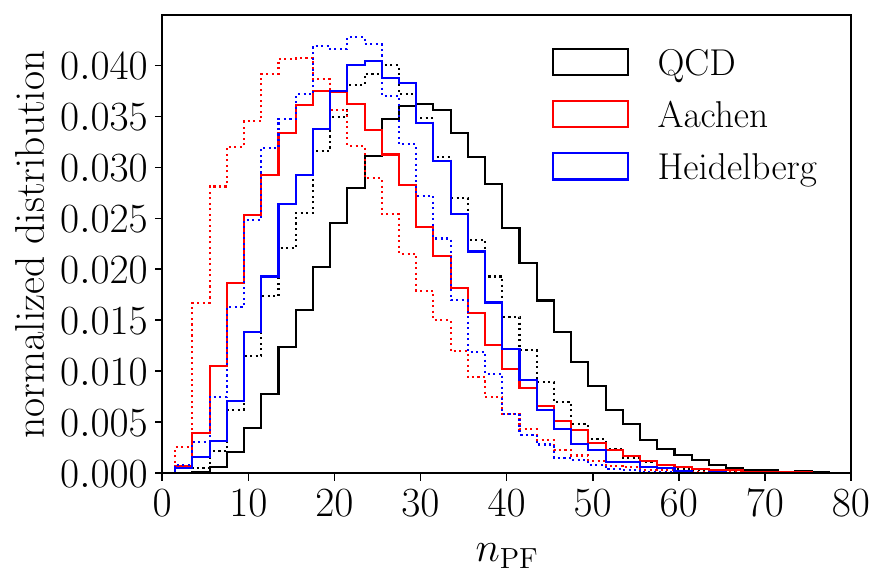}
  \includegraphics[width=0.33\textwidth]{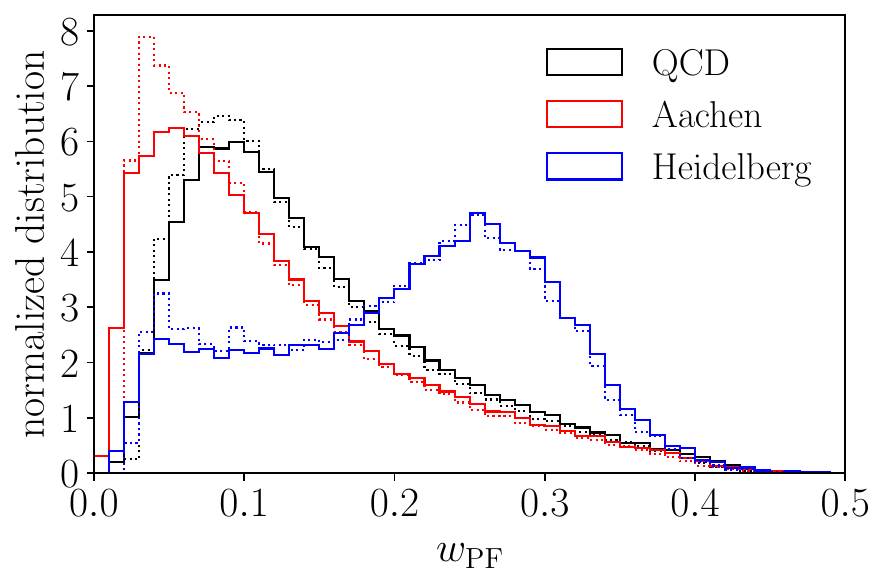} 
  \includegraphics[width=0.33\textwidth]{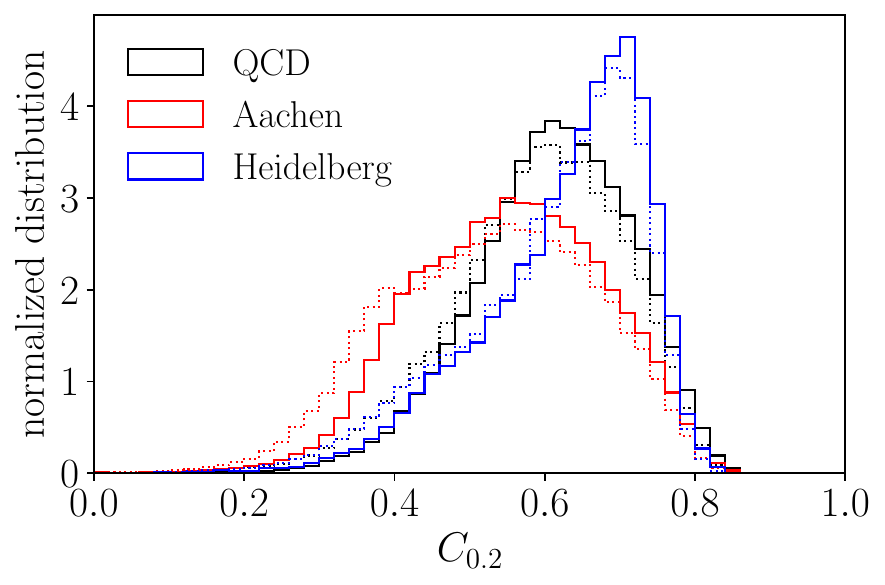}
  \includegraphics[width=0.33\textwidth]{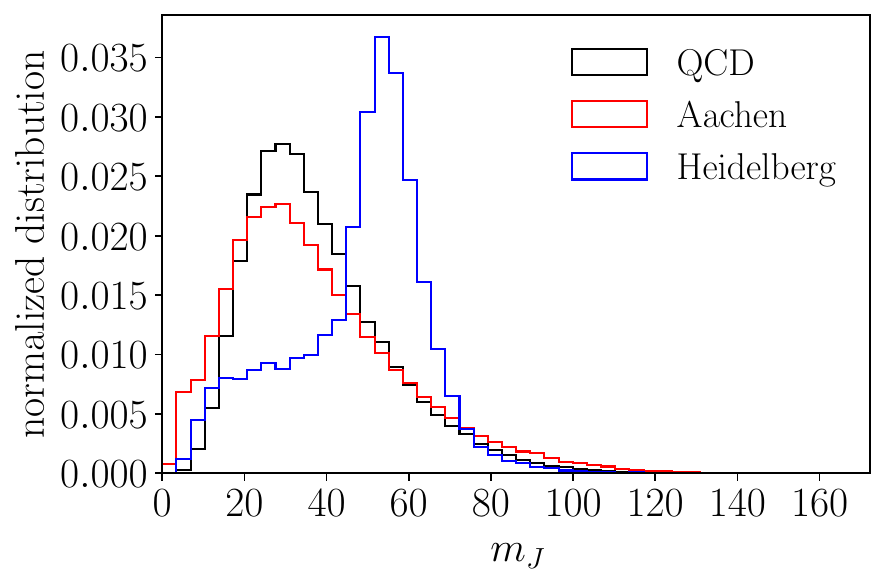}
  \includegraphics[width=0.33\textwidth]{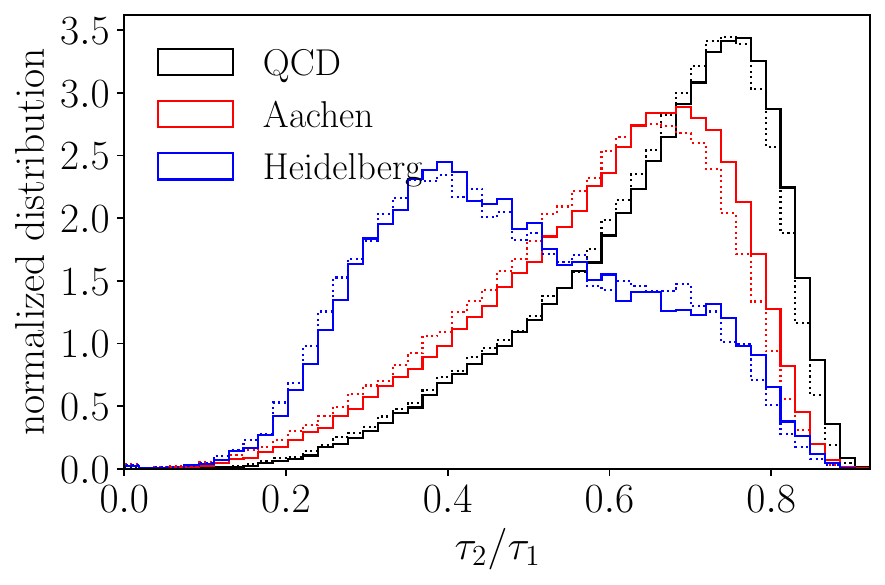}
  \includegraphics[width=0.33\textwidth]{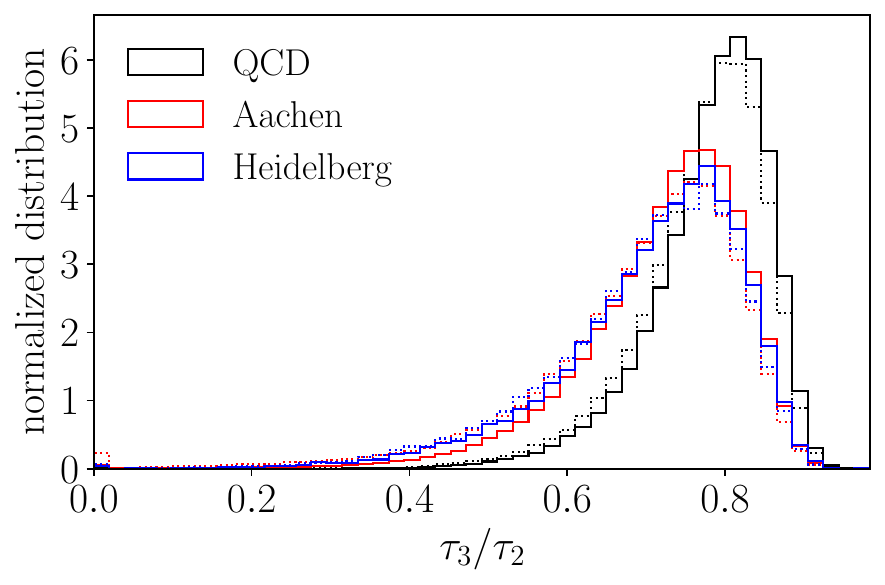}
  \caption{High-level observables calculated from the four-vectors of
    the constituents (solid) and the calorimeter pixels (dotted)
    comparing the QCD jets with the Aachen and Heidelberg dark jets.}
  \label{fig:obs}
\end{figure}

To illustrate the physics behind the QCD and signal jets, we can use
standard subjet observables. First, we count the number of jet
constituents ($n_\text{PF}$) and compute its radiation distribution or
girth ($w_\text{PF}$)~\cite{Gallicchio:2010dq}, which can be
reconstructed from particle-flow output or from jet images.  In
addition, the two-point energy correlator $C_{0.2}$ is known to
separate for instance quark and gluon jets~\cite{Larkoski:2013eya},
\begin{align}
n_\text{PF} =  \sum_i 1\;, \qqquad
w_\text{PF} = \frac{\sum_i p_{T,i}  R_{i,\text{jet}}}{ \sum_i p_{T,i}} \;, \qqquad
C_{0.2} &= \frac{\sum_{ij} p_{T,i} p_{T,j} (R_{ij})^{0.2} }{\left(\sum_i p_{T,i}\right)^2} \; .
\label{eq:qg_obs}
\end{align}
We show these observables in Fig.~\ref{fig:obs}, both calculated from
the jet constituents and from the pixelized jet image. While the
number of soft constituents is similar for all three samples, with
slightly smaller numbers for the semi-visible jets, the Heidelberg
dark jets show clear signs of massive decays and separated
prongs. Obviously, the same pattern is visible in the jet mass, which
in this case reflects our choice of $m_{q_d} = 50$~GeV. Because
preprocessing of the jet images includes normalizing the pixel
entries, we can extract the jet mass only from the constituents.  The
shoulder towards smaller jet masses corresponds to jets where we lose
hard constituents.

Finally, we can track the number of prongs through the IR-safe
N-subjettiness ratios~\cite{Thaler:2010tr}, which should agree between
constituent-based and image-based definitions. Only the Heidelberg
dataset shows a significant deviation from the democratic limit,
$\tau_i/\tau_j \simeq 1$, in the ratio $\tau_2 /\tau_1 \simeq 0.4$,
indicating a 2-prong structure in a fraction of its dark jets. The
slight difference between the constituent-based and image-based
high-level observables is easily explained by the finite resolution of
the jet images and the lack of IR-safety for the number of
constituents. The jet mass and the N-subjettiness show the smallest
effects, as one would expect by construction. We also emphasize that
a sensitivity to soft and collinear splittings is
only problematic when we compare high-level observables to
perturbative QCD predictions, but not when we train supervised or
unsupervised classification networks on simulations or data.

\subsection{Energy flow polynomials}
\label{sec:data_efp}

\begin{figure}[t]
  \includegraphics[width=0.33\textwidth]{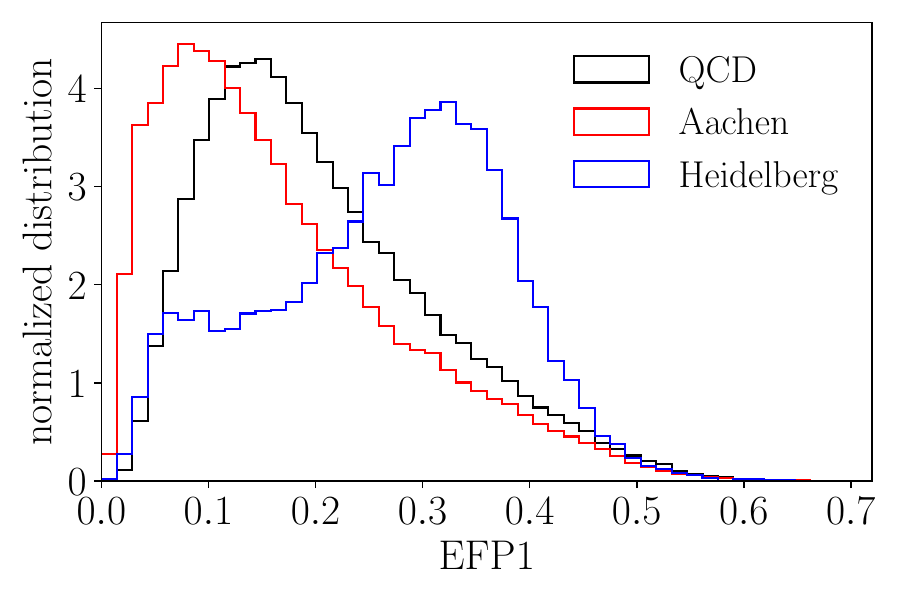}
  \includegraphics[width=0.33\textwidth]{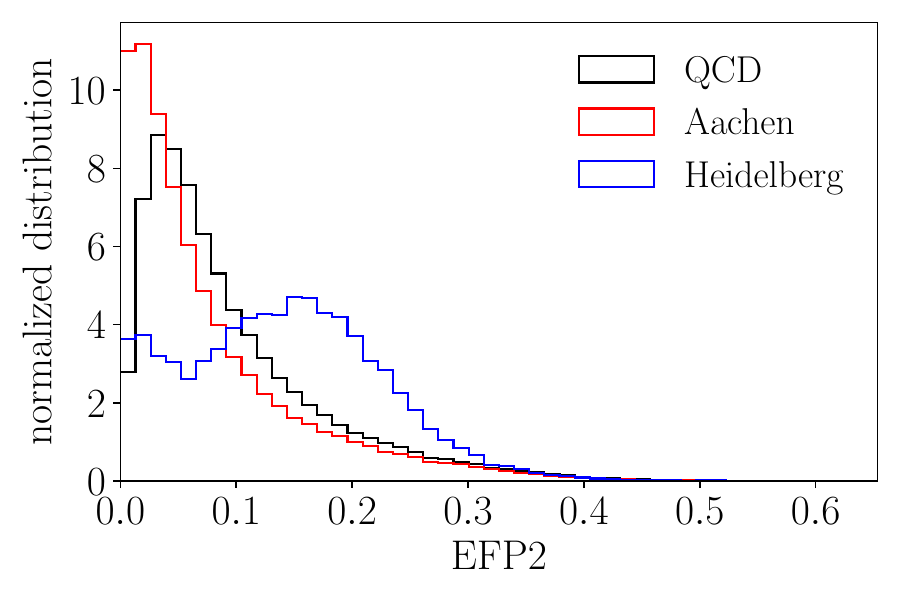}
  \includegraphics[width=0.33\textwidth]{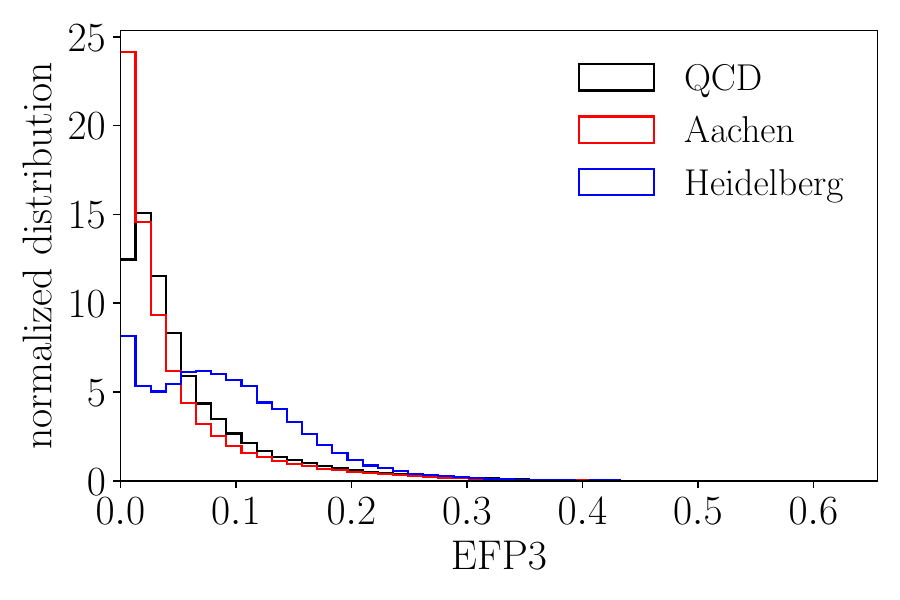}
  \includegraphics[width=0.33\textwidth]{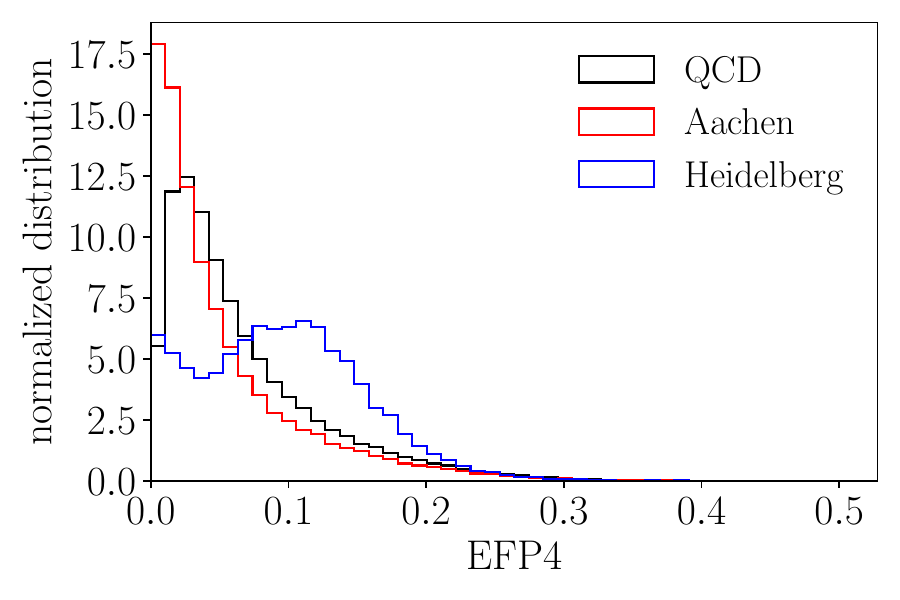}
  \includegraphics[width=0.33\textwidth]{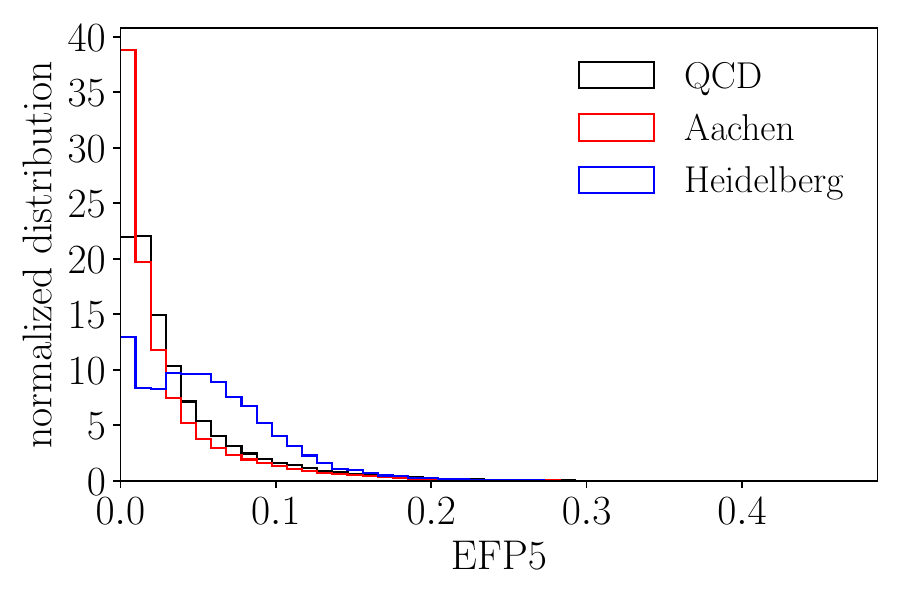}
  \includegraphics[width=0.33\textwidth]{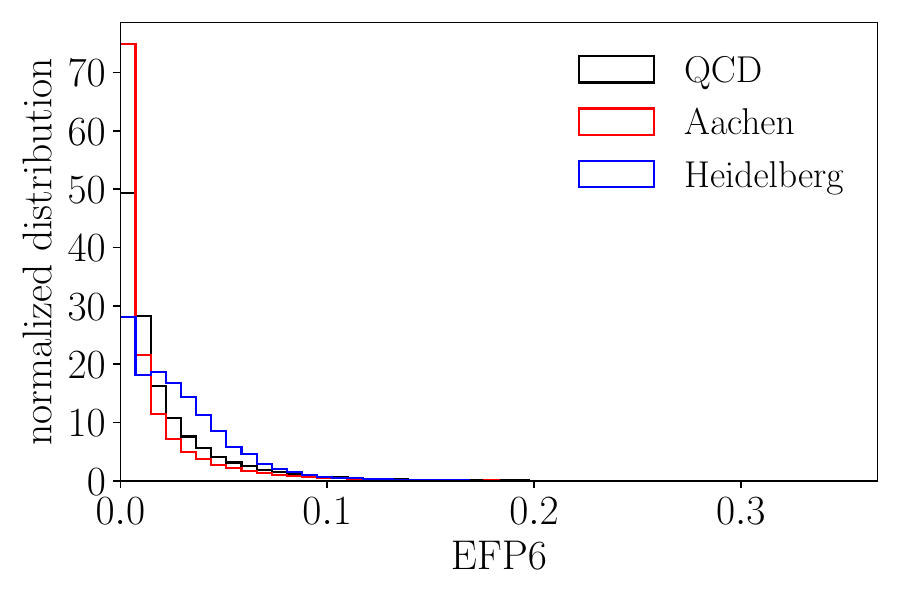}
  \caption{Leading EFPs for QCD jets and for the Aachen and Heidelberg
    dark jets, computed from the jet constituents.}
  \label{fig:efp_1d}
\end{figure}

Energy flow polynomials (EFPs)~\cite{Komiske:2017aww} provide a
powerful systematic description of the phase space patterns of jets,
described by the constituents' transverse momenta and their
geometric separation,
\begin{align}
  z_i = \frac{p_{T,i}}{p_{T,J}}\;,
  \qqqquad
  R_{ij} = \sqrt{ (\Delta y_{ij})^2+ (\Delta \phi_{ij})^2} \; .
\end{align}
We use prime EFPs, which are not a product of EFPs with lower rank,
and remove the constant EFP.  For a maximum of order three in the angular
distance $R_{ij}$ this leaves us with the eight EFPs
\begin{align}
  \text{EFP1} &= \sum_{i_1,i_2} z_{i_1} z_{i_2} \; R_{i_1 i_2}  &
  \text{EFP5} &= \sum_{i_1,i_2,i_3} z_{i_1} z_{i_2} z_{i_3} \; R_{i_1 i_2} R_{i_1 i_2} R_{i_1 i_3} \notag \\
  \text{EFP2} &= \sum_{i_1,i_2} z_{i_1} z_{i_2} \; R_{i_1 i_2} R_{i_1 i_2} \approx 2\frac{m_J^2}{p_{T,J}^2} &
  \text{EFP6} &= \sum_{i_1,i_2,i_3} z_{i_1} z_{i_2} z_{i_3} \; R_{i_1 i_2} R_{i_2 i_3} R_{i_1 i_3} \notag \\
  \text{EFP3} &= \sum_{i_1,i_2} z_{i_1} z_{i_2} \; R_{i_1 i_2} R_{i_1 i_2} R_{i_1 i_2} &
  \text{EFP7} &= \sum_{i_1,i_2,i_3,i_4} z_{i_1} z_{i_2} z_{i_3} z_{i_4} \; R_{i_1 i_2} R_{i_1 i_3} R_{i_1 i_4} \nonumber \\
  \text{EFP4} &= \sum_{i_1,i_2,i_3} z_{i_1} z_{i_2} z_{i_3} \; R_{i_1 i_2} R_{i_1 i_3}  & \qqquad
  \text{EFP8} &= \sum_{i_1,i_2,i_3,i_4} z_{i_1} z_{i_2} z_{i_3} z_{i_4} \; R_{i_1 i_2} R_{i_2 i_3} R_{i_3 i_4} \; .
\end{align}
Depending on the application, we can reweight the relative importance
of the momentum fraction and the angular separation by replacing
\begin{align}
z_i \to z_i^\kappa
\qquad \text{and} \qquad
R_{ij} \to R_{ij}^\beta \; ,
\label{eq:def_kappabeta}
\end{align}
with appropriate values for $\kappa$ and $\beta$. In
Fig.~\ref{fig:efp_1d} we show the first six EFPs for the signal and
background datasets. As for the high-level observable, we see a clear
difference between the QCD and Aachen datasets on the one hand and the
Heidelberg dataset on the other. This difference is linked to the
two-prong structure and the finite jet mass of this signal sample. We
can understand the signal pattern in the Heidelberg sample for EFP2,
where from the threshold for the jet kinematics in
Eq.\eqref{eq:kinematics} we can estimate
\begin{align}
  \frac{m_J}{p_{T,J}}
  \lesssim \frac{50~\gev}{150~\gev} 
  \qquad \Rightarrow \qquad
  \text{EFP2} \lesssim \frac{2}{9} \; .
\end{align}
The fact that the same pattern appears in almost all EFPs indicates that the EFPs are strongly correlated.

\subsection{Preprocessing}
\label{sec:data_pre}

\begin{figure}[t]
  \includegraphics[width=0.95\textwidth]{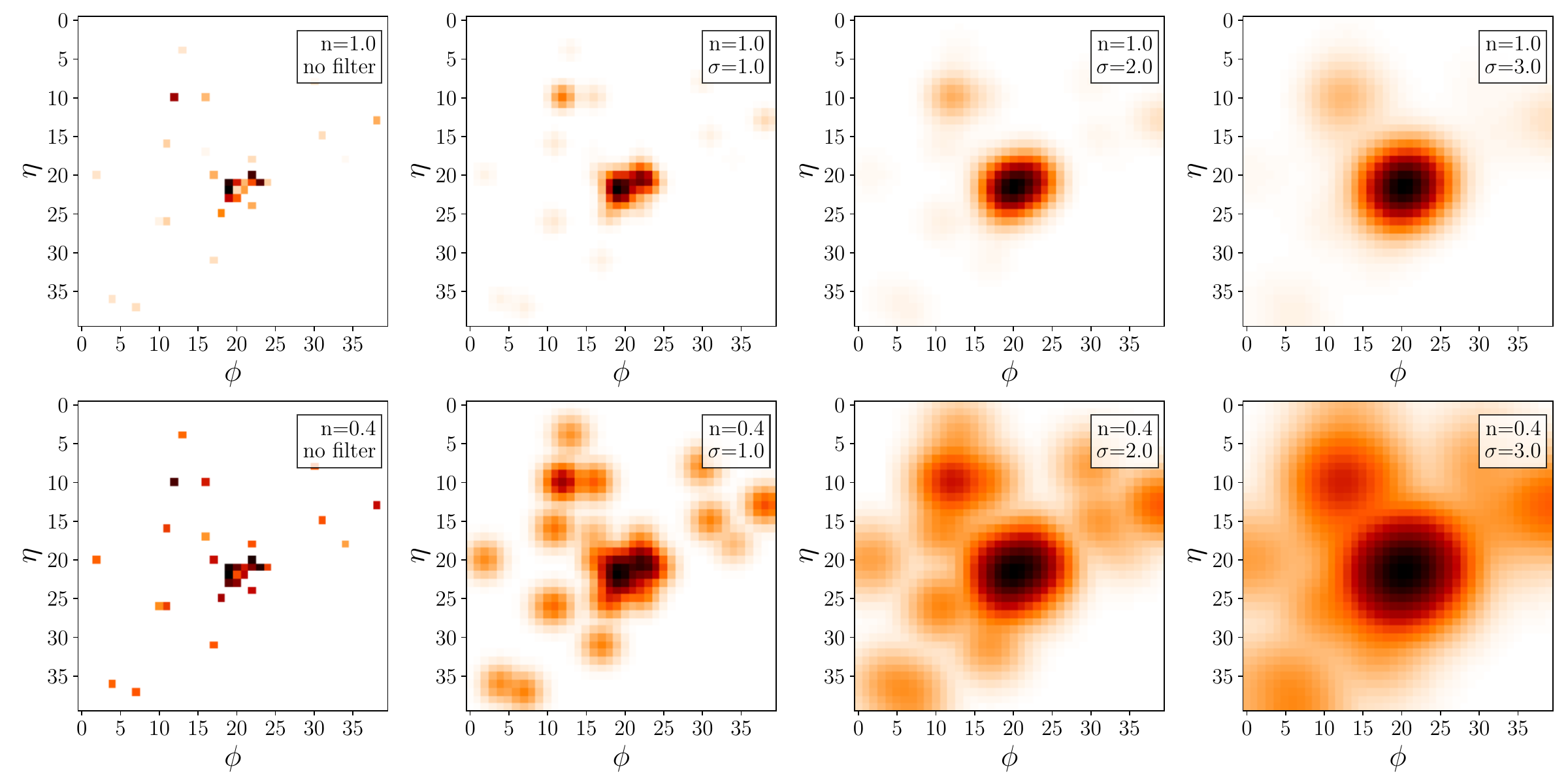}
  \caption{Illustration of two different $p_T$ reweightings (top and
    bottom) and the Gaussian filter with a varying width $\sigma$,
    applied to a single QCD jet image.}
  \label{fig:gfilter_remap}
\end{figure}

For our analysis we can apply various preprocessing steps to the jet
images, some of them already outlined in
Ref.~\cite{deOliveira:2015xxd}, and some of them with the specific
goal of enhancing the sensitivity of a given anomaly detection
method. In general, we also expect such preprocessings to affect the
sensitivity to specific physics
signals~\cite{Baldi:2020hjm,Canelli:2021aps}\footnote{We note also
  that the preprocessing could be replaced by learning representations
  that are invariant to symmetry transformations and augmentations of
  the data using self-supervised contrastive learning, as in JetCLR
  \cite{Dillon:2021gag}.}.

The first choice, crucial for any neural network, is how we define the
$p_T$-information per jet constituent. A naive jet image
representation typically uses the sum of the constituent $p_T$ as the
pixel value. Both Standard Model and anomalous features in jets occur
at very different $p_T$-scales; hard decays lead to features at $p_T
\gg 1$~GeV, while jets with a modified parton showering are sensitive
to GeV-scale constituents, similar to quark-gluon tagging. This means
the standard choice biases a classification or anomaly detection
technique towards features at high $p_T$~\cite{Baldi:2020hjm,Finke:2021sdf}.  This
explains why autoencoders tag jets with higher complexity more easily
if complexity or structure is usually assumed to affect the harder
features of the jet.  However, the two signal datasets in this paper
fall into two different categories, the Heidelberg sample being more
and the Aachen sample being less complex than the QCD background.

For our preprocessing we start by defining a dimensionless input and
normalize each image such that the total intensity summed over all
pixels is one.  For typical loss functions, compressing a wide range
of (input) values leads to an improved numerical performance.  To
study and exploit a potential bias, we consider different choices to
the pixel intensity in the jet image,
\begin{align}
  p_T \; \rightarrow \; p_T^n
  \qquad \text{with} \qquad n \in (0,1] \;. 
\label{eq:scalings}
\end{align}
Established working points include square root reweighting $(n=1/2)$ or
$n=1/4$~\cite{Finke:2021sdf}.  Aside from the obvious choice of the
pixel transverse momentum, the alternative reweightings stretch the
resolution at low transverse momenta and move the peak of the pixel
distribution to higher reweighted intensity values. This allows the
network to extract more information from the large number of soft
pixels, while keeping most of the information on hard, decay related
pixels. 
The reweightings also change the density of the jets in physics space, and given that we
define anomalous jets as those in the low density regions, they physically alter what jets are anomalous.
For an optimal network training it might eventually be
beneficial to provide the network with two reweighted inputs, one
focusing on soft pixels and one focusing on hard
pixels~\cite{Baldi:2020hjm}.

Second, we need to deal with the sparsity of the jet images. We use a
Gaussian filter to decrease the sparsity, convoluting the image with a
Gaussian smearing kernel with the width $\sigma=0.5~...~3.0$
pixels. This filter also correlates neighbouring pixels. In
Fig.~\ref{fig:gfilter_remap} we illustrate the effect of the
$p_T$-reweighting and the Gaussian filter on a single QCD jet image.
The smearing is equivalent to using a correspondingly defined kernel
MSE loss~\cite{Finke:2021sdf} and provides a better measure of
similarity for jet images than employing the standard mean squared
error (MSE) distance measure. If the intensity in a bright pixel is
shifted to a neighbouring pixel, the resulting image is closer to the
original image than an image where the intensity is moved further
away.
Note that neither the reweighting of the pixel intensity nor the application of the Gaussian kernel destroys information.
The reweighting is done with a bijective mapping. The Gaussian kernel applied is fixed and contains no randomness.
It is therefore invertible except for edge effects where intensity can be smeared out of the image.
However, we expect these effects to be negligible, as the image range in $\Delta\eta$ and $\Delta\Phi$ is sufficiently large compared to the jet radius.
\medskip

\begin{figure}[t]
  \includegraphics[width=0.33\textwidth]{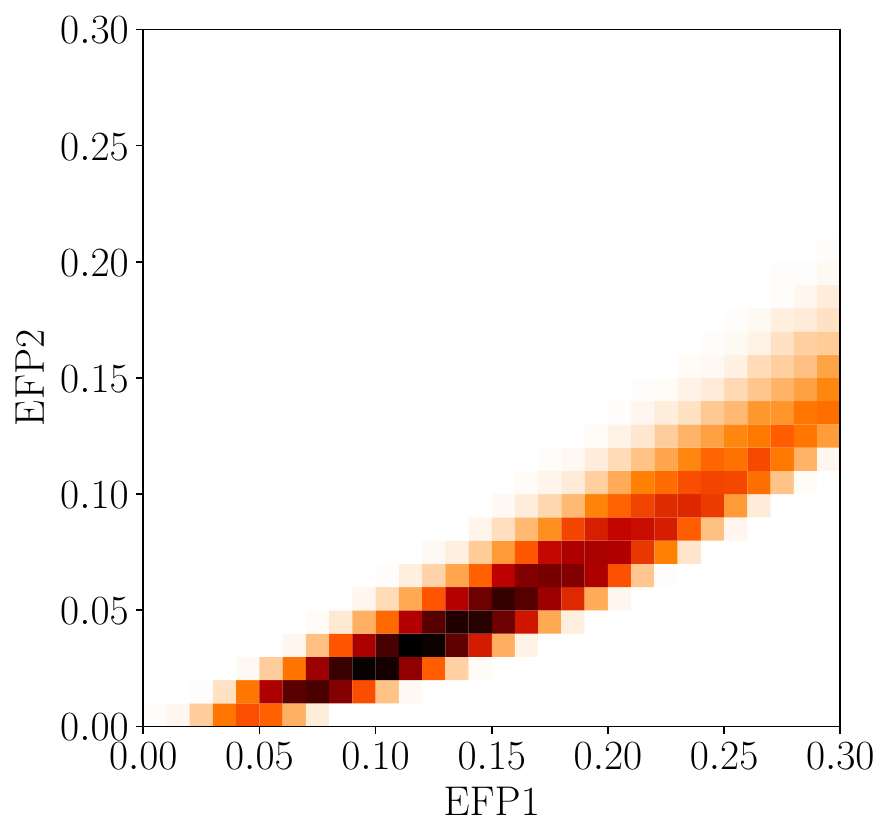}
  \includegraphics[width=0.33\textwidth]{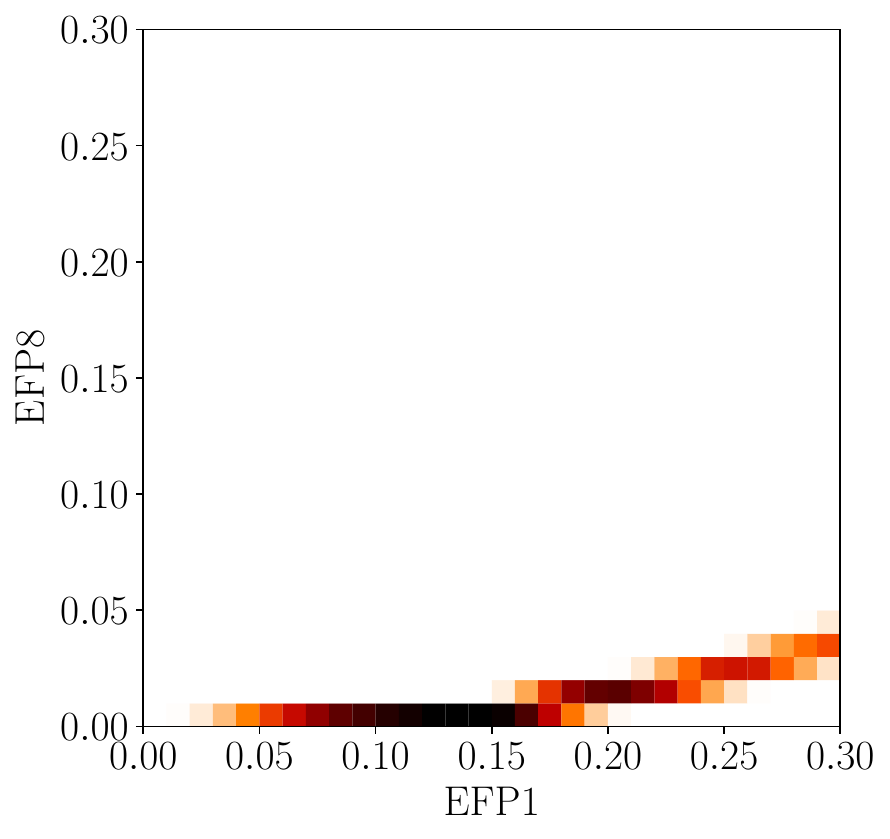}
  \includegraphics[width=0.33\textwidth]{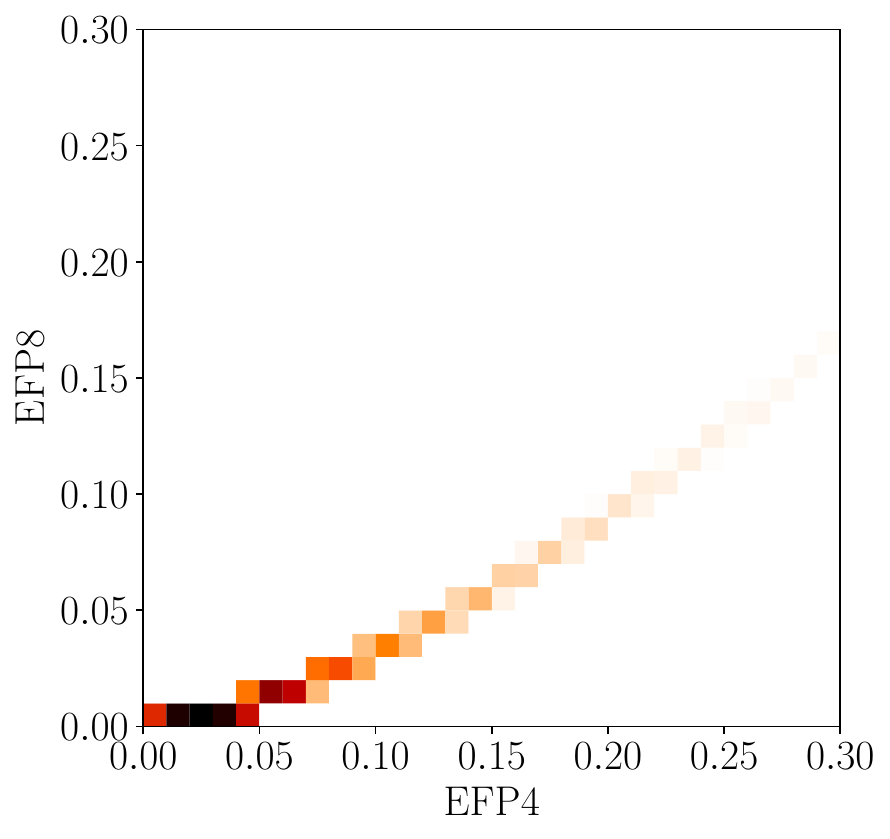}
  \includegraphics[width=0.33\textwidth]{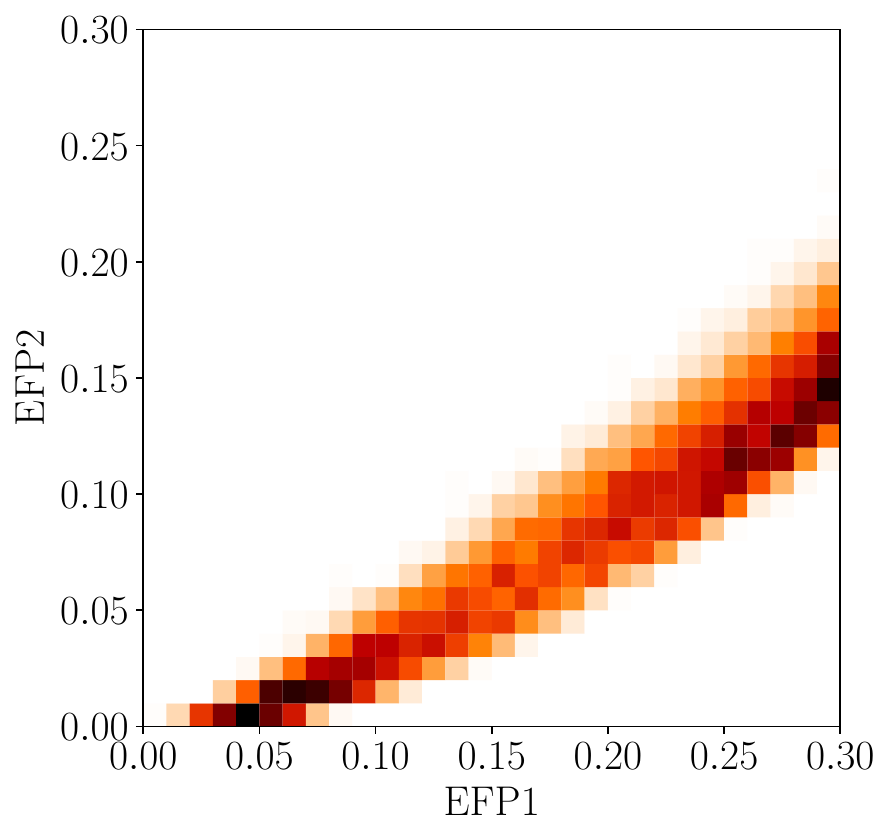}
  \includegraphics[width=0.33\textwidth]{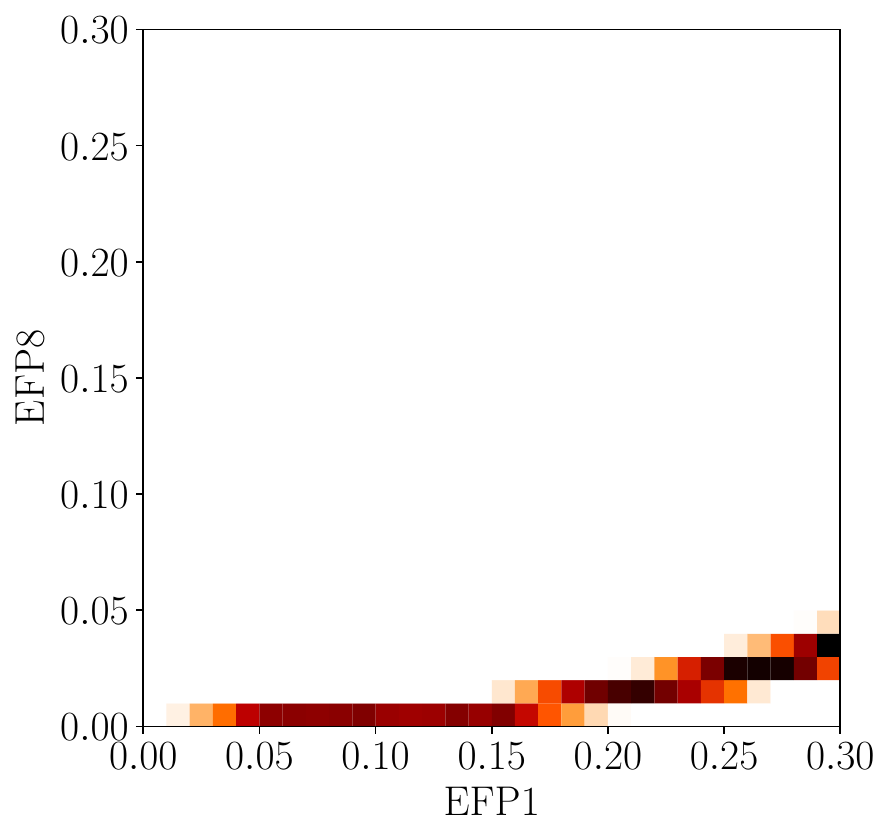}
  \includegraphics[width=0.33\textwidth]{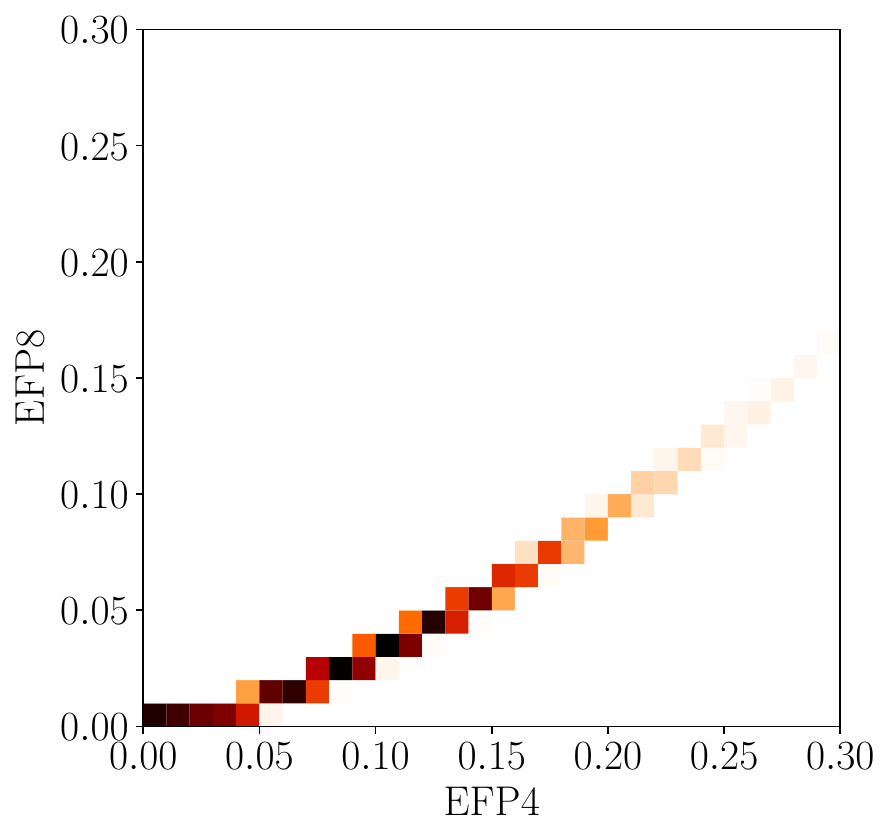}
  \caption{Sample correlations among some of the EFPs for QCD jets (upper) and
    the Heidelberg dark jets (lower).}
  \label{fig:efp_2d}
\end{figure}

For networks working on EFPs we can apply similar preprocessing steps,
either reweighting the momentum fractions $z_i$ individually or
reweighting the EFPs as a whole.  To exploit some of the analytic
properties of the INN we choose the latter option, including a
reweighting $\text{EFP}_i \to \log \text{EFP}_i$. The Gaussian filter
is not relevant for the EFP representation. To estimate the
correlations between the different EFPs we turn to
Fig.~\ref{fig:efp_2d}.  Indeed, the EFPs we use as network input are
strongly correlated in a non-linear way.  Because of these
correlations, for instance the QCD jets and the Heidelberg dark jets
have a similar overall structure in the 8-dimensional EFP space, but
populate different parts of the sub-manifolds due to the one-prong vs
two-prong difference.

Given these strong correlations, we can either train the network to
extract the relevant information after extracting the correlations, or
we can provide the network with a decorrelated input constructed from
the first eight EFPs. To stabilize the training and to save training
time we choose the second option and use principle component analysis
(PCA). For this we first subtract the mean of the distribution from
each data point. We then change to the eigenvalue base of the
covariance matrix, removing all linear correlations between the
individual components. Finally, we scale the individual components so
that their standard deviation is one. The resulting distribution is as
close to a normal distribution as is possible using a linear
transformation.

\section{K-means}
\label{sec:kmeans}

Anomaly searches are not limited to deep learning. One can also employ
a variety of classical ML methods, in particular for density
estimation. However, density estimation for multidimensional data,
like jet images, is notoriously difficult. Standard approaches like
kernel methods or density estimation based on histograms scale badly
with the data dimensions $d$ and the number of training points $n$.

For anomaly detection, it is not crucial to know the actual
density. Any anomaly score which is (strongly) correlated with the
density can potentially be useful. We use the well known k-means 
clustering algorithm to define such anomaly scores. Loyd's k-means algorithm \cite{Lloyd} scales linearly with the number of data points and dimensions for each iteration. Since it usually converges quickly \cite{informationretrieval} (in our application $\sim300$ iterations are sufficient for convergence), we can apply it to large datasets with high-dimensional data. K-means provides a given number $k$ of clusters with centroids
\begin{align}
  \vec{\mu}_i=\frac{1}{N_i} \sum_j \vec{r}_{i,j} \; ,
\end{align}
where the vectors $\vec{r}_{i,j}$ represent the data instances $j$
assigned to cluster $i$, and $N_i$ is the number of data instances in
cluster $i$. The clusters divide the data into a patchwork of Voronoi
cells. We use the Lloyd's k-means algorithm with 10 different initializations of the centroids following the "k-means++" prescription~\cite{kmeans} and pick the one with the lowest inertia, as implemented
in the scikit-learn python library~\cite{scikit-learn}.

K-means is neither a density estimation nor an anomaly detection
algorithm. However, assigning an effective size to each cluster $i$
around its centroid, e.g.
\begin{align}
  \rho_i=\frac{1}{N_i} \sum_j |\vec{r}_{i,j}-\vec{\mu}_i| \; ,
\end{align}
(with $j$ iterating over the vectors assigned to cluster $i$) the clusters map out the large-scale data distribution, and we can
construct several useful anomaly scores. In this context, using many
clusters seems to be beneficial to approximate the underlying
distribution as precisely and detailed as possible. On the other hand,
the number of data points in each cluster has to be large enough to
assign a statistically meaningful size. We employ k-means with $k=100$
clusters. Using 100000 training images, the smallest cluster contains
78 jet images, and all others more than 100 jet images.  All our
models are trained using $40\times40$ jet images that are convoluted
with a Gaussian smearing kernel with $\sigma=3$ pixels.

In addition to the datasets described in Sec.~\ref{sec:data}, we also
test the anomaly-detection methods discussed in this section on the
standard benchmark set for top tagging~\cite{Kasieczka_top}. It
consists of QCD and top jets with $p_T = 550~...~650$~GeV, in contrast
to the low $p_T = 150~...~300$~GeV QCD jets used for our dark
jets. The jet constituents are processed into images in the same way
as described in Sec.~\ref{sec:data}.  Following the discussion in
Refs.~\cite{Dillon:2021nxw,Finke:2021sdf}, we take QCD jets as a
background and top jets as the anomalous signal (direct top tagging)
as well as top jets as a background and QCD jets as the signal
(reverse top tagging). These additional tagging examples help to
illustrate the differences between different anomaly scores based on
k-means.

\subsection{K-nearest centroids}
\label{sec:data_knc}

A simple anomaly score for a jet image, which does not take into
account the cluster sizes, is the minimal distance to one of the
k-means cluster centroids. We refer to this anomaly score as MinD. A
similar approach was discussed in Ref.~\cite{Fraser:2021lxm},
where instead of k-means clustering a k-medoids algorithm was used to
obtain the representatives of the background dataset. The MinD score
assumes that regular datapoints are close to the k-means centroids,
whereas outliers are not. However, MinD has several obvious
drawbacks. K-means itself is susceptible to outliers far from the main
distribution, since it may assign a cluster to a single outlier or a
small group of outliers. Obviously, the number of clusters is a
crucial parameter in this context. Moreover, points on the boundary
between two clusters have a higher anomaly score than points close to
a cluster center, even if both clusters are part of smooth
distributions. For our dataset, we expect such a smooth distribution
rather than a collection of well separated clusters.

Both problems can be mitigated by using a score based on $k$ nearest
neighbors. In the standard method the distance of a point to its $k$
nearest neighbors is used to estimate the probability density at this
point or to determine its affiliation with a class of points.  Here,
we do not consider the distance to the $k$ nearest data instances but
to the $k$ nearest cluster centers (KNC).  We define the anomaly score
KNC5 as the average distance of a data point to the $k=5$ nearest
centroids obtained through k-means clustering.

\begin{figure}[t]
  \includegraphics[width=0.45\textwidth]{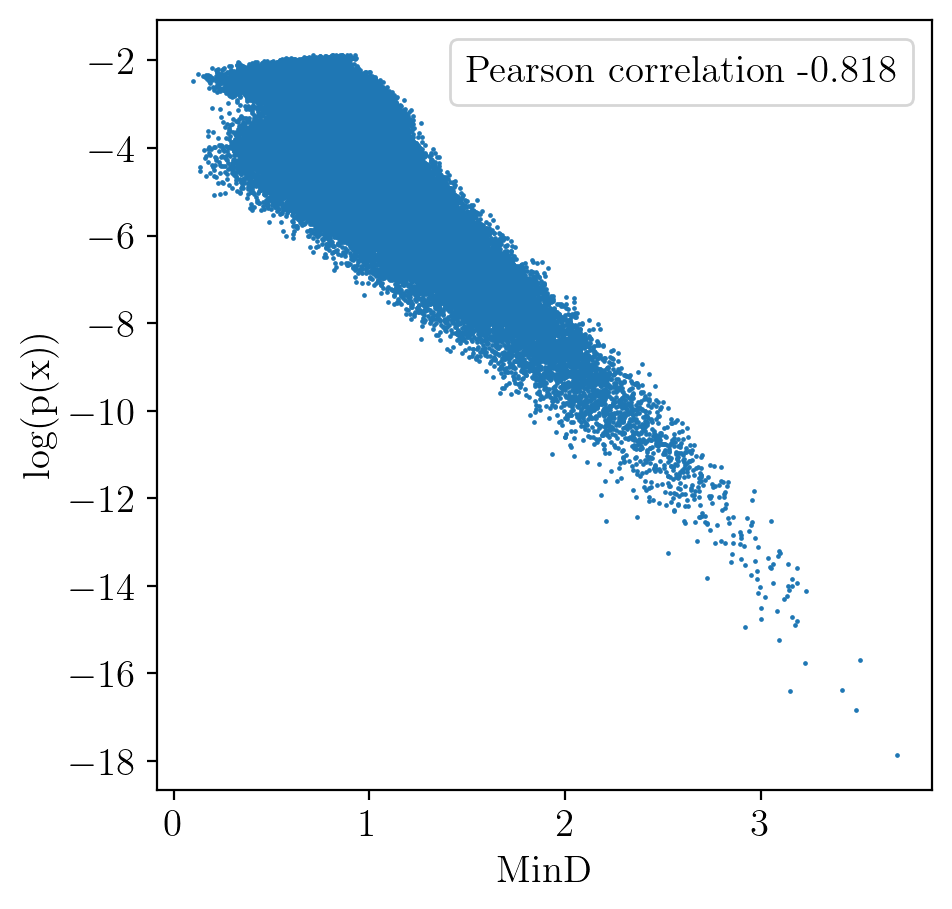}
  \hspace*{0.07\textwidth}
  \includegraphics[width=0.45\textwidth]{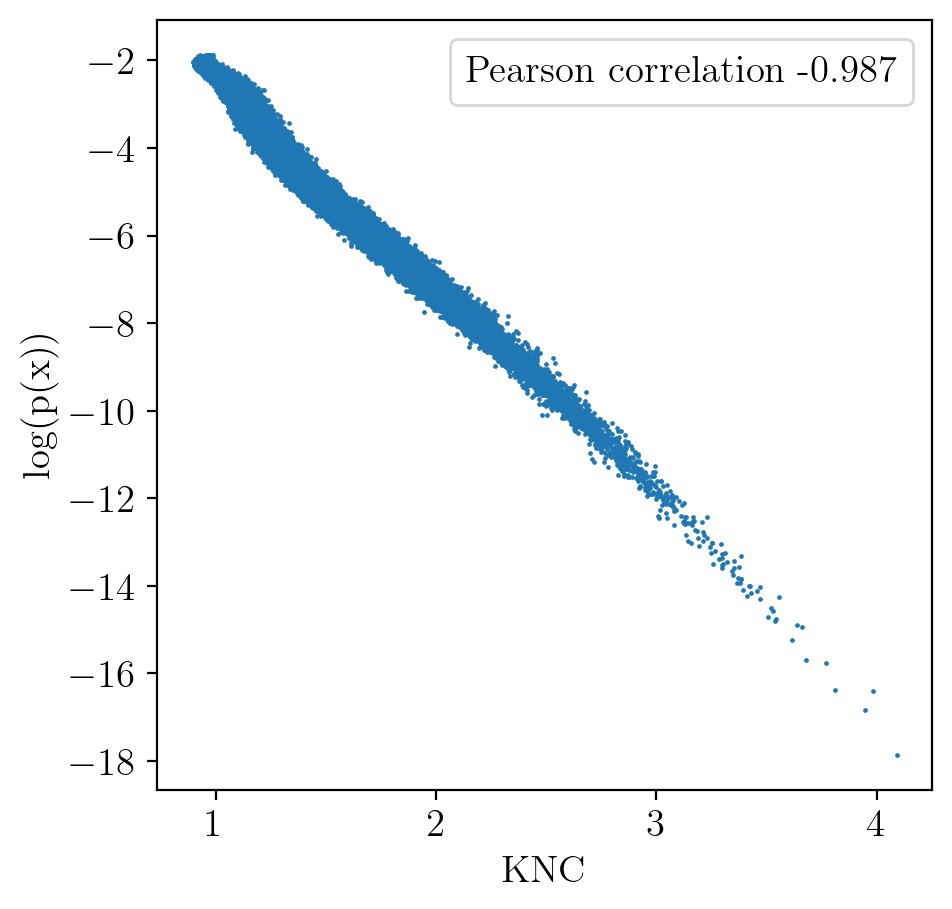}
  \caption{The logarithmic density as a function of MinD (left) and
    KNC5 (right) for a data set drawn from a five-dimensional normal
    distribution with unit covariance. All 100000 training points are
    shown.}
	\label{fig:correlation}
\end{figure}

Fig.~\ref{fig:correlation} shows MinD and KNC5 for 100 k-means
clusters on a data set with 100000 training points drawn from a
five-dimensional normal distribution with unit covariance. We observe
a strong correlation between MinD and the logarithm of the density. As
expected, KNC5 improves the correlation significantly. Although this
simple example might not be representative for more complex data
distributions, we expect that KNC5 provides a better correlation of
the anomaly score and the density than MinD for most smooth
distributions.

\subsection{Gaussian mixture model}
\label{sec:data_gmm}

As a benchmark for our k-means method we use a Gaussian mixture model
(GMM), a standard density estimation technique where the density is
approximated by a sum of several multidimensional Gaussians. The GMM
uses an iterative expectation maximization algorithm to fit the
Gaussians to the data distribution. On the one hand, the GMM provides
another simple benchmark.  On the other hand, a GMM is an obvious
generalization of k-means based algorithms since a GMM provides a mean
for each of the mixture components. These means are equivalent to the
k-means centroids if one uses an expectation maximization with a
covariance matrix proportional to the unit matrix and a common
variance, which approaches zero~\cite{bishop:2006:PRML}.  In this
limit, each data point is assigned to one of the mixture components as
it is assigned to one cluster in k-means.

We use the scikit-learn python library~\cite{scikit-learn} to fit a
GMM to our jet images smeared with our standard Gaussian kernel with
$\sigma = 3$. Usually all entries of the covariance matrix in a GMM
are fit parameters. However, the 1600 dimensions of the jet images,
with a $1600\times1600$ covariance matrix, are prohibitive for the
full fit. Instead, we use spherically symmetric Gaussians with a
variance $\alpha_i+\beta$ for each mixture $i$, where $\alpha_i$ is a
fit parameter and $\beta$ is a regularization parameter. We decrease
$\beta$ until the smallest fitted variance reaches $10^{3} \times
\beta$. This way, we ensure that $\beta$ does not dominate the
variance, as it would be the case if we used the default scikit-learn
parameter $\beta=10^{-6}$.

For smooth distributions one would expect the $\sqrt{\alpha_i}$ to be of the
same order of magnitude as the typical length scale of the dataset,
e.g.\ the $\rho_i$ of the k-means clusters (see
Tab.~\ref{tab:dims}). However, the GMM finds $\sqrt{\alpha_i}$ which are
roughly two orders of magnitude smaller. This mismatch is due to the
fact that the effective dimension of our data is much smaller than
1600, implying the data lives in a lower-dimensional subspace. Fitting
a spherical Gaussian to the data that has almost zero variance in many
of its dimensions will result in a strongly underestimated standard
deviation. Hence, the actual likelihood estimation is very
poor. Nevertheless, using the negative log-likelihood of the GMM to
define the GMMLL anomaly score might still be valuable. The insight
regarding the effective dimension of the data distribution motivates a
new density-based anomaly score using k-means clustering, introduced
next.

\subsection{Likelihood-inspired anomaly scores}
\label{sec:data_like}

Building a regularly shaped histogram to estimate the density of our
1600-dimensional data space is of course impossible due to the curse
of dimensionality. Instead, we propose to use the k-means clusters as
generalized bins which are automatically adapted to the underlying
distribution, in analogy to the Gaussians centered around their means
in the GMM.

To be specific, we approximate each cluster as a multidimensional
sphere with radius $\rho_i$ around its centroid $\vec{\mu}_i$. In the
spirit of density estimation, we associate a likelihood to find a data
instance $\vec{r}$ associated to cluster $i$ according to
\begin{align} 
  \mathrm{L}_i(\vec{r})
  = \mathcal{N}_i \; 
\begin{cases}
  1  & \mbox{for}\quad |\vec{r}-\vec{\mu}_i| < \rho_i \\[2mm]
  \left(\dfrac{\rho_i}{|\vec{r}|}\right)^{d-1}
  \exp \left[ -\dfrac{(|\vec{r}-\vec{\mu}_i|-\rho_i)^2}{2\sigma_i^2} \right] & \mbox{for}\quad |\vec{r}-\vec{\mu}_i|>\rho_i \\
\end{cases} \notag \\[3mm]
\text{with} \quad
\sigma_i^2=\frac{1}{N_i} \sum_j (|\vec{r}_{i,j}-\vec{\mu}_i|-\rho_i)^2\; .
\label{eq:pdf_formula}
\end{align}  
with $j$ iterating over vectors assigned to cluster $i$.
Here, $\mathcal{N}_i$ is a normalization factor and $d$ an effective
dimension to be discussed below. Inside the cluster we assume a constant density. However,
the clusters are not taken as spheres with sharp boundaries, but we
add Gaussian tails such that outliers have different scores depending
on the distance to the cluster border. The tails also ensure that the
likelihoods of points in the gap between two close, but not
overlapping clusters can add up. The factor in front of the Gaussian
is chosen such that the marginal one-dimensional likelihood
$\mathrm{L}_i(|\vec{r}_{i,j}-\vec{\mu}_i|)$ resembles the observed
distribution of the data instances. The factor is proportional to the
inverse of the volume factor connecting $\mathrm{L}_i(\vec{r})$ and
$\mathrm{L}_i(|\vec{r}_{i,j}-\vec{\mu}_i|)$ in a space with dimension
$d$.

\begin{figure}[b!]
  \centering
  \includegraphics[width=0.75\textwidth]{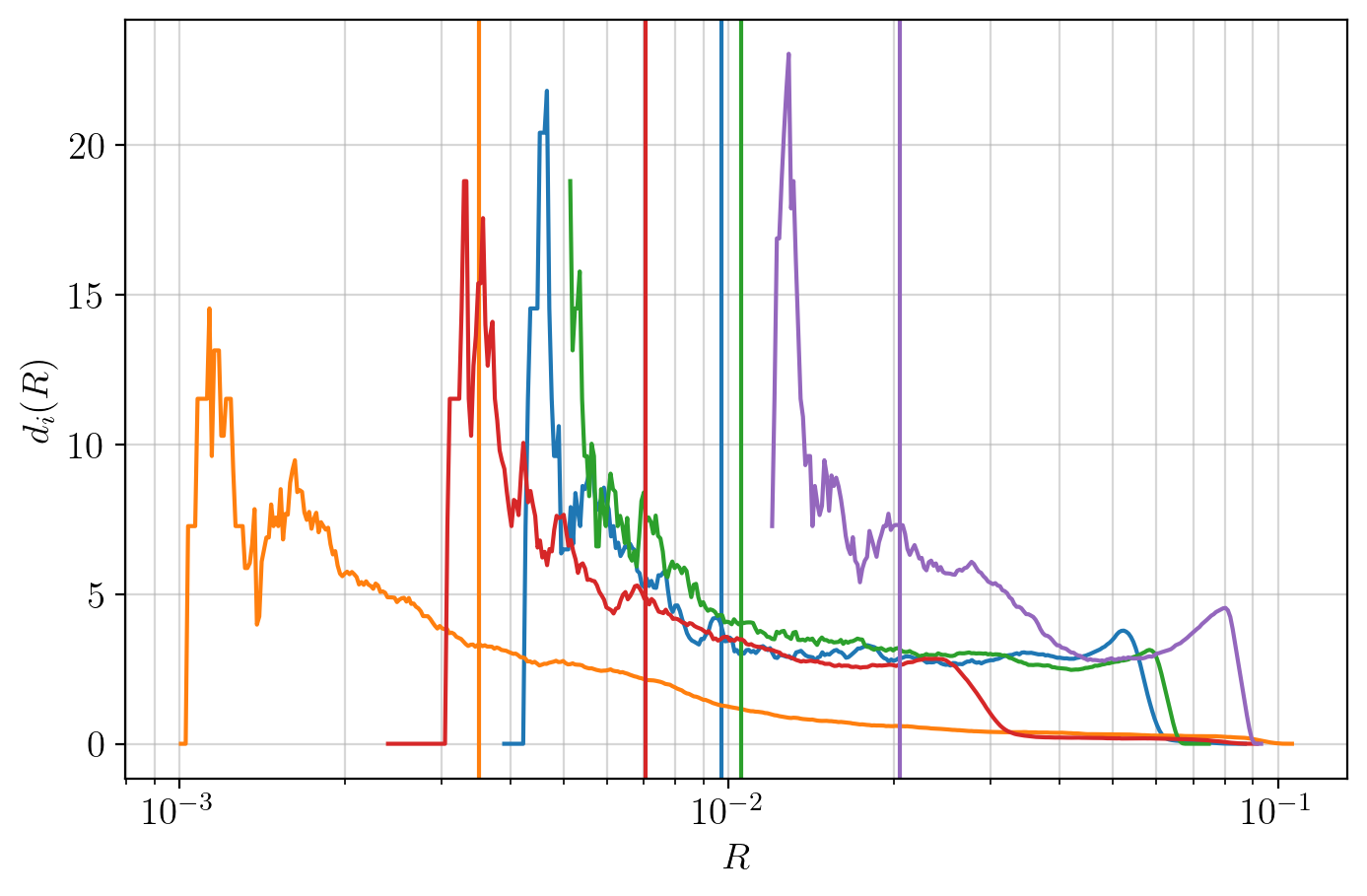}
  \caption{Effective dimensionality $d(R)$ for five different
    high-$p_T$ QCD clusters. The vertical lines are located at
    $R=\rho_i$ for each cluster. The intersection point is the
    estimate for the effective dimension of the
    cluster.}
  \label{fig:dims}
\end{figure}

For density estimation based on histograms, the normalization is
defined as $\mathcal{N}_i= N_i/V_i$, where $V_i$ denotes the volume of
the bin. In our multidimensional problem it is non-trivial to estimate
the volume $V_i$. One option is $V_i\propto \rho_i^{1600}$. As a
consequence, two clusters with only slightly different radii
$\rho_{i}\neq \rho_{j}$ would have extremely different densities
inside the clusters. However, in Sec.~\ref{sec:data_gmm}, we have
already discussed that the effective dimensions of the subspace where
the data lives is much smaller.

For a sensible normalization, one has to approximately determine the
effective number of dimensions.  Making the simplifying assumption
that the underlying density $n$ of data points in and around a cluster
is uniform, a sphere of radius $R$ in $d$ dimensions contains $N(R)=n
\pi^{d/2} R^d/\Gamma(d/2+1)$ data points. This equation can be solved
for the dimensionality of the dataset $d_i(R)=\ln (N_i(R)/N_i(c
R))/\ln(c)$ at a scale $R$, where $N_i(R)$ is the number of training
points in the sphere of radius $R$ around the cluster centroid, and
$c$ is a scaling factor which we set to $c=1.1$. The relevant length
scale for each cluster is assumed to be $\rho_i$, such that the
effective dimensionality of a cluster is defined as
$d_i(\rho_i)$. Fig.~\ref{fig:dims} shows $d_i(R)$ for five clusters
of the high-$p_T$ QCD images of the top-tagging dataset together with
the corresponding $\rho_i$ values. The effective dimension is
estimated in a region with sufficient statistics, where the curve is
rather smooth. This means we have chosen $c$ sufficiently large for
sufficient statistics, but also small enough to access the shape of
$d_i(R)$. We notice that as $R$ approaches the size of the actual
distribution as a whole, its dimensionality decreases to zero, since
any distribution looks point-like from a large distance. As shown in
Tab.~\ref{tab:dims}, the median values for the effective
dimensionality of the k-means clusters range between 5 and 8,
depending on the data preprocessing. The same value
$d=\text{med\;} (d_i(\rho_i))$ is then used to calculate $\mathrm{L}_i(\vec{r})$
for all clusters.\medskip

Our anomaly tagging algorithm can be summarized as:
\begin{center}
\framebox{\parbox{14cm}{
\begin{enumerate}
	\item{Perform k-means clustering on a dataset, using $k=100$.}
	\item{Compute $\rho_i$, $\sigma_i$, and $d_i=d(\rho_i)$ for each cluster $i$ from the distribution of points inside cluster $i$.}
	\item{Find $d=\text{med\;}(d_i)$ as a representative effective dimension for all clusters.}
	\item{Compute the normalization factor $\mathcal{N}_i$ by requiring $\int_{\mathcal{R}^d}\mathrm{L}_i(\vec{r})d^d\vec{r} = N_i$, 
	i.e.\ the likelihood is integrated in $d$ dimensions.
}
	\item{For each data point $\vec{r}$ compute $\mathrm{L}_i(\vec{r})$ for each cluster as defined in Eq.\eqref{eq:pdf_formula}.}
	\item{Compute the anomaly score $-\log(\mathrm{L}(\vec{r}))$ with $L(\vec{r})=\sum_{i=0}^k \mathrm{L}_i(\vec{r})$.
	 }
\end{enumerate}
}}
\end{center}
The corresponding anomaly score is called MLLED (k-Means based
Log-Likelihood estimation in Effective Dimensions).\medskip

\begin{table}[b!] 
  \centering
  \begin{small} \begin{tabular}{c|ccccc} 
\toprule
&$\min \rho_i $&$\text{med\,} \rho_i$ &$\max \rho_i $&$\max \dfrac{\sigma_i}{\rho_i} $& $\text{med\,} d_i(\rho_i)$\\ 
\midrule
high $p_T$ QCD & 0.0031  & 0.011 & 0.026  & 0.32 & 5.2\\ 
high $p_T$ top & 0.0127  & 0.015 & 0.026  & 0.27 & 5.4\\ 
low $p_T$ QCD  & 0.0040  & 0.012 & 0.022  & 0.28 & 5.8\\ 
low $p_T$ QCD with $\sqrt[4]{p_T}$& 0.0088 & 0.012 & 0.018 & 0.25 & 7.8 \\ 
\bottomrule
\end{tabular} \end{small}
\caption{Properties of the set of 100 clusters found by k-means for
  the different datasets used as background for anomaly tagging.}
\label{tab:dims}
\end{table} 

For the QCD jets in our background datasets, we find cluster radii
$r_i$ which vary by up to one order of magnitude, i.e.\ we find a
variation of likelihoods by a factor of roughly $10^5$ using the
estimated effective dimensions. A point in one of the smallest
clusters has a density roughly $10^5$ larger than a point in a cluster
which is 10 times larger. Hence, a data point has to be a few sigmas
away from the small cluster to have the same small likelihood as a
data point in the large cluster. The higher the dimensionality of the
clusters the more we put weight on assigning high anomaly scores or
low likelihoods to the points in the low-density clusters, as compared
to the points that are outliers of the small and highly populated
clusters. If the dimensionality is too high, the dependence of the
score on the cluster size will dominate over the dependence on the
distance to the cluster border. Hence, out-of-cluster anomalies that
might lie only a few sigmas away from the border cannot be
distinguished anymore from probably non-anomalous points inside
smaller clusters. Accordingly, the tagging performance for such
anomalies is strongly diminished. As we will see, this is the case for
tagging dark jets from the Aachen dataset or QCD jets in a top-jet
background. On the other hand, some types of anomalies reside in the
low but non-zero background density region of the data space. In these
cases the correct hierarchy of clusters and thus a significant
dependence on the cluster size and population is required. Such
examples may include the tagging of top jets in a QCD background
(reverse top tagging).

According to the previous discussion, different normalization factors
in Eq.\eqref{eq:pdf_formula} may lead to anomaly scores being more or
less sensitive to different types of anomalies. Hence, we also
consider an anomaly score based on the unrealistic assumption
$d=1$. This is no longer an estimate for the density, but might still
perform well as an anomaly score to tag out-of-cluster outliers with
good precision while preserving a certain hierarchy for the in-cluster
densities. We refer to this anomaly score as MLL1D.

For k-means the radius of a cluster $\rho_i$ is anti-correlated with
the number of data points $N_i$ in the cluster. Taking only $\rho_i$
or $N_i$ into account for the normalization may thus suffice for a
qualitatively correct ordering of their densities. Using $d=1$ but
$\mathcal{N}_i=N_i$ instead of the choice defined above defines the
MLLN anomaly score.

\subsection{Performance}
\label{sec:kmeans_results}

\begin{figure}[t]
	\includegraphics[width=0.49\textwidth]{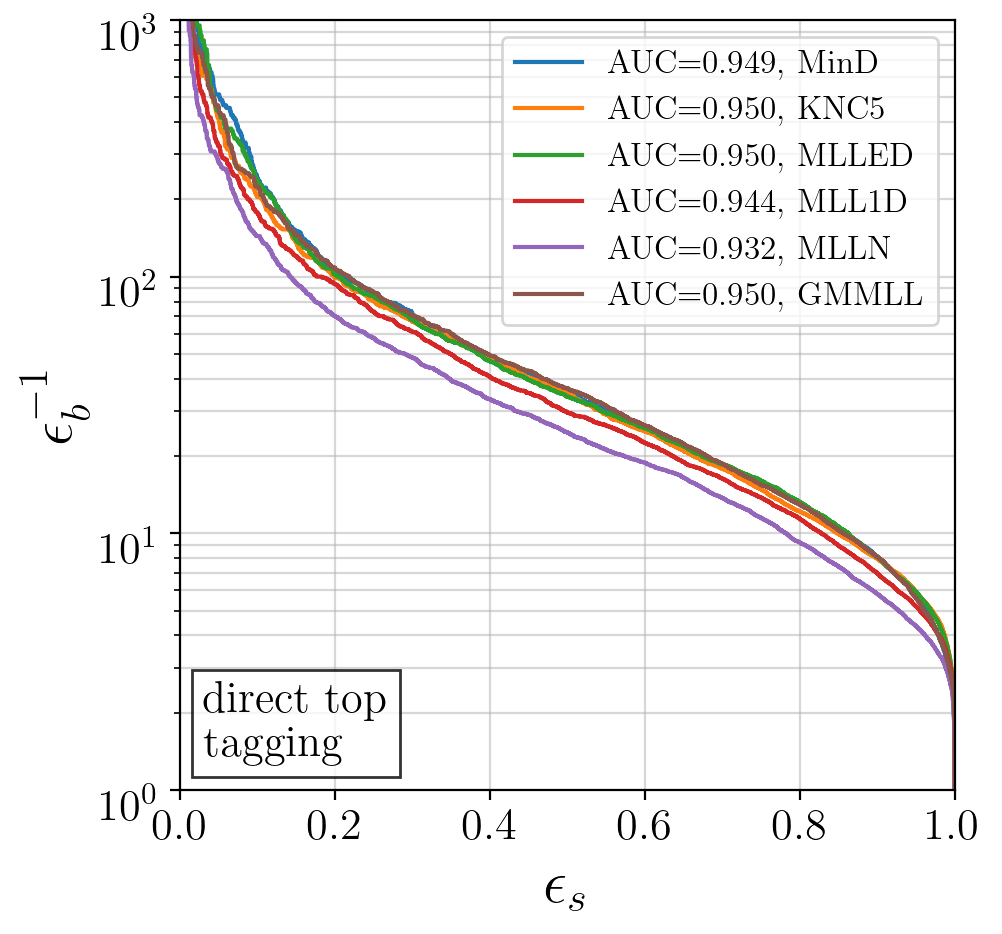}
	\includegraphics[width=0.49\textwidth]{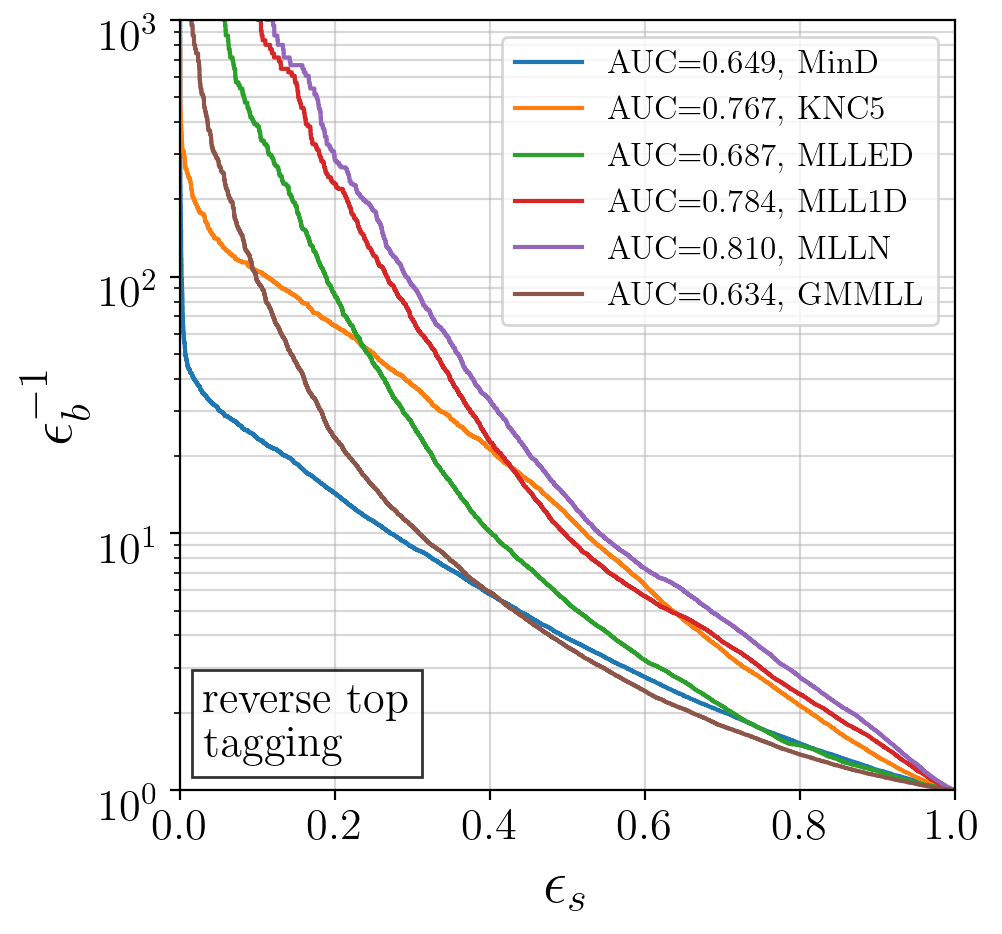}
	\caption{ROC curves for tagging high-$p_T$ top jets in a QCD
		background (left) and high-$p_T$ QCD jets in a top background
		(right). The various anomaly scores are discussed in the text.}
	\label{fig:ROC_knc_top}
\end{figure}

\begin{figure}[t]
	\includegraphics[width=0.49\textwidth]{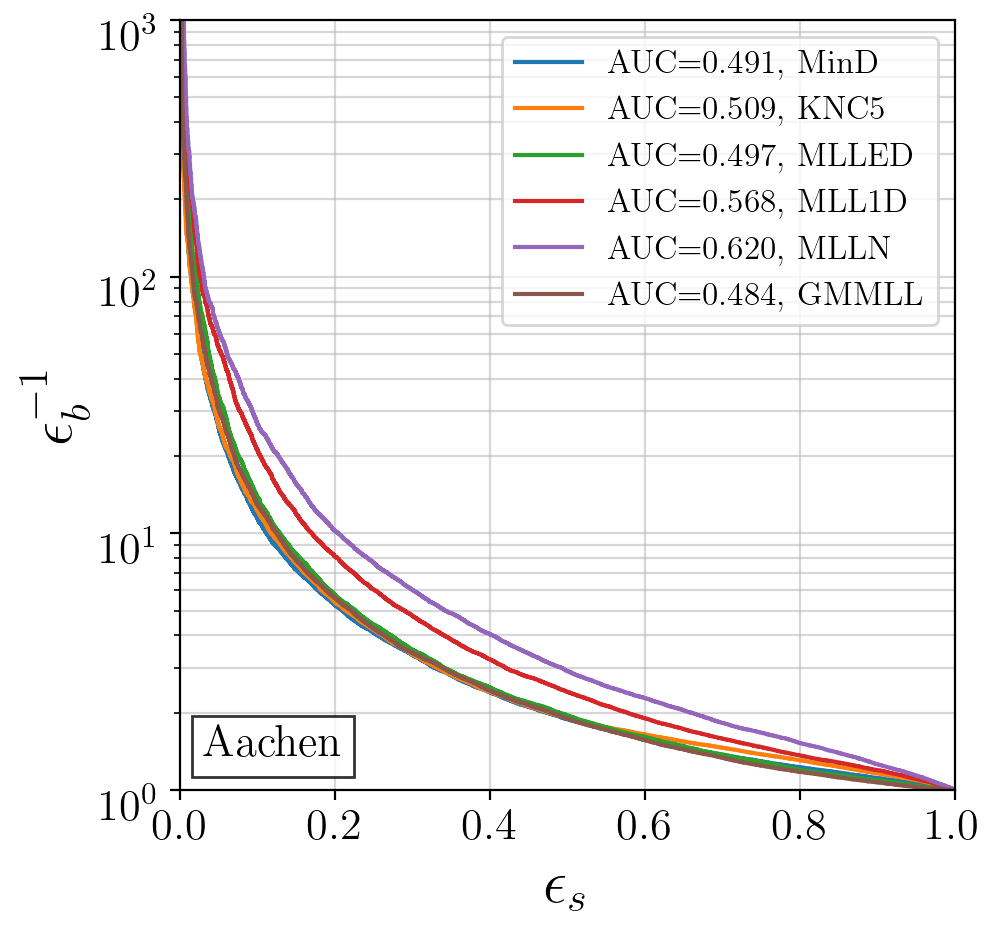}
	\includegraphics[width=0.49\textwidth]{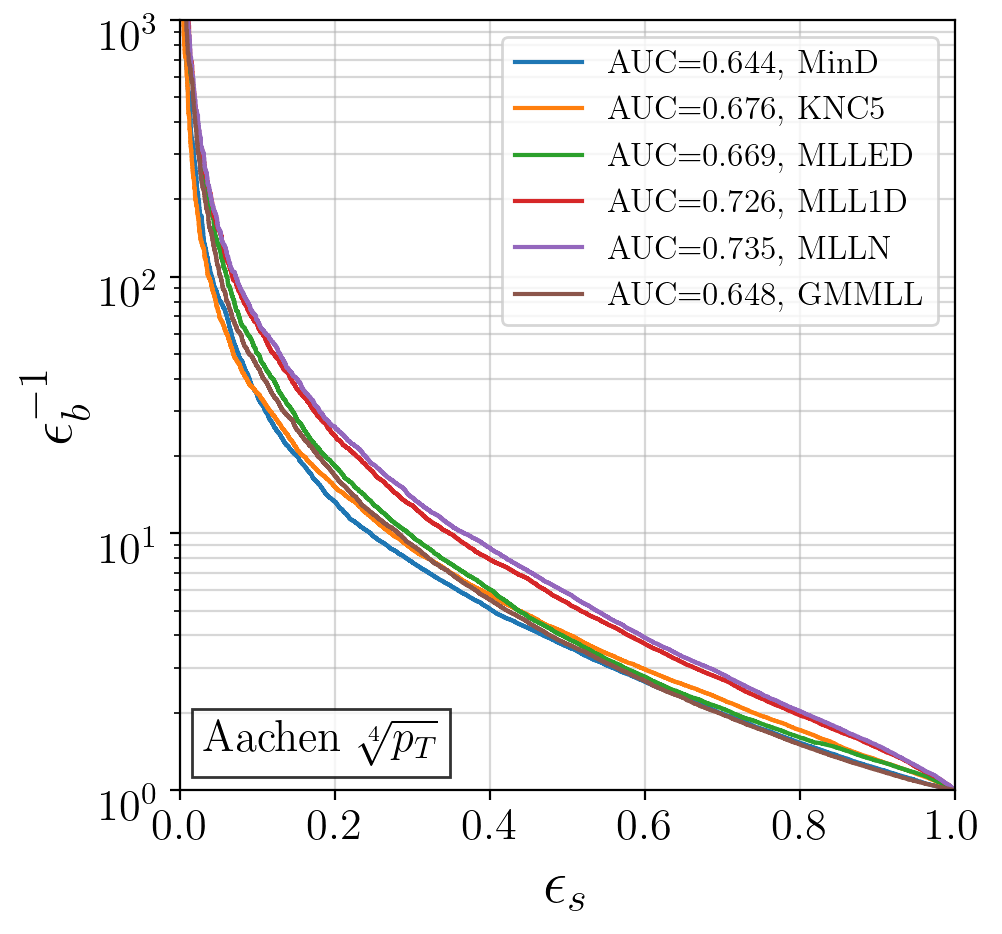}
	\includegraphics[width=0.49\textwidth]{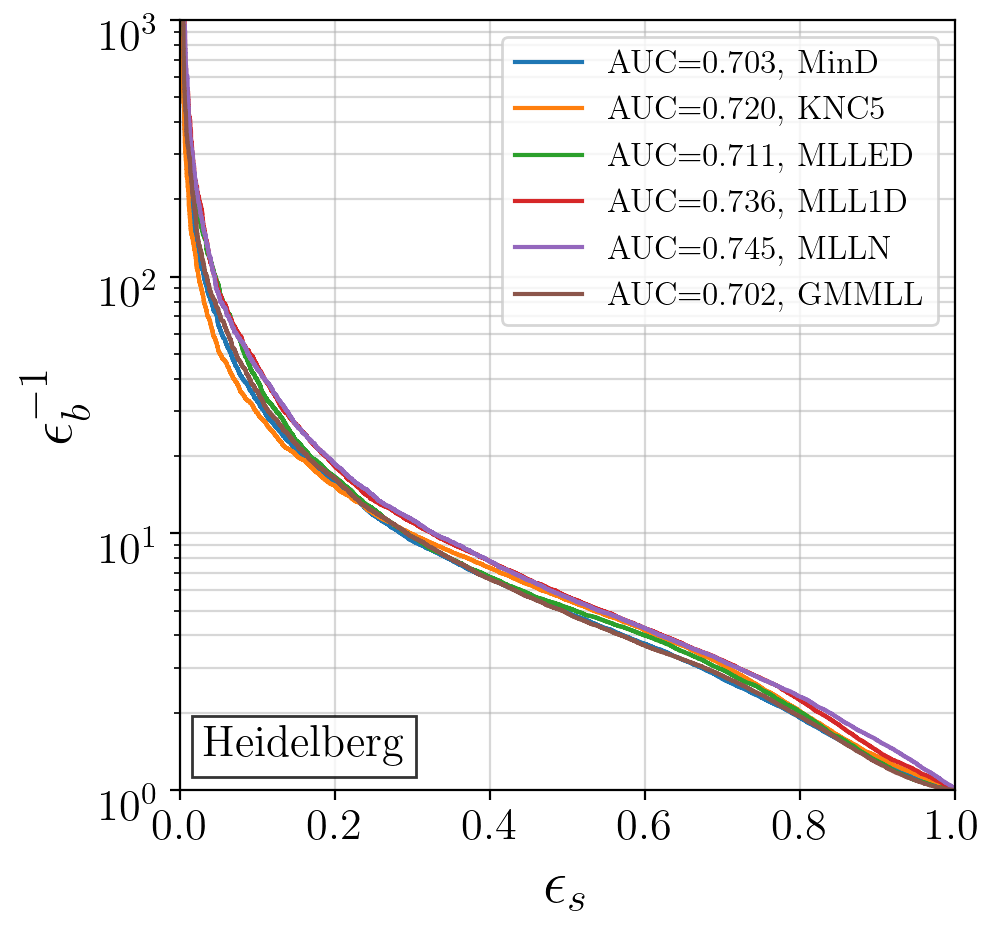}
	\includegraphics[width=0.49\textwidth]{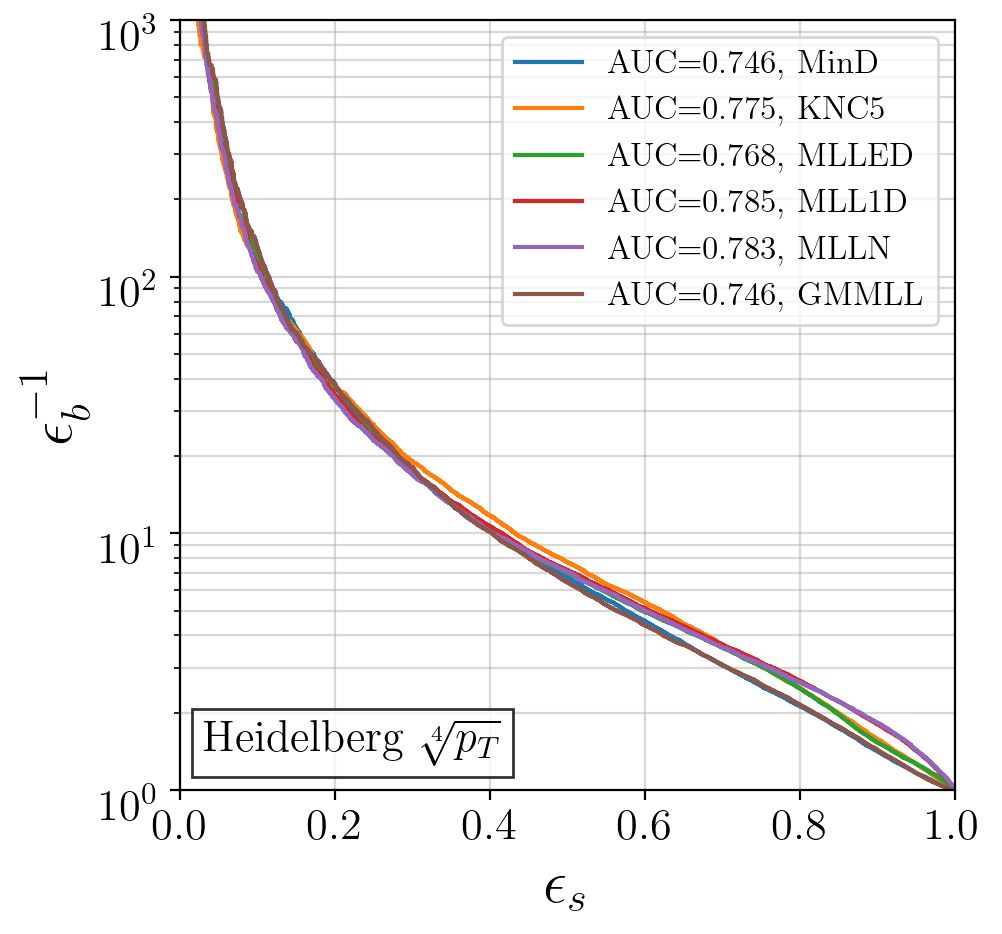}
	\caption{ROC curves for tagging Aachen (top row) and Heidelberg
		(bottom row) dark jets in a QCD background. The pixel intensities
		in the images are given by the $p_T$ within a pixel (left) or
		$p_T^{1/4}$ (right). The various anomaly scores are discussed in
		the text.}
	\label{fig:ROC_knc}
\end{figure}

After defining the different k-means anomaly scores, we discuss their
performance on our benchmark datasets. Top jets have a prominent
3-prong structure and do not require special preprocessing in order to
highlight the features relevant for top-tagging in a QCD
background. On the other hand, QCD jets and dark jets are dominated by
the intensity in the central pixels. In these cases it is promising to
apply preprocessing to highlight the dim features, for instance using
a $p_T^{1/4}$ reweighting. It is evident from Tab.~\ref{tab:dims}
that such a preprocessing also increases the number of effective
dimensions of the clusters, as the number of relevant pixels
increases.

The ROC curves for direct and reverse top tagging are shown in
Fig.~\ref{fig:ROC_knc_top}, and for tagging dark-jet anomalies in
Fig.~\ref{fig:ROC_knc}. We see that MinD, KNC5, MLLED and MGG all have
similar performances, except for the reverse top tagging
case. Ignoring the cluster size completely in assigning a density,
MLLN has a low performance on the top-signal dataset, but it performs
best on the Aachen dataset and for reverse top tagging. The reason for
this, as discussed at the end of Sec..~\ref{sec:data_like}, is that
MLLED and MLLN are sensitive to different kinds of anomalies. MLL1D is
a compromise between MLLED and MLLN. It does not show the best, but a
reliable performance on all five tasks. This renders MLL1D the most
model-agnostic anomaly detection algorithm in this section.

In line with what we find for the anomaly detection methods presented
in the following chapters, we observe a strong improvement in the dark
jet tagging performance after preprocessing the jet images.  The
Gaussian filter combined with the 4th root reweighting are essential
for the applicability of the MSE distance measure on which our
clustering and density estimation algorithms rely. Using $p_T$ instead
of $p_T^{1/4}$ implies neglecting the contribution of the low-$p_T$
pixels in the distance measure between images, whereas performing no
filtering will result in a distance being artificially large for
images with a small spatial shift of intensity.

\section{Dirichlet-VAE}
\label{sec:dvae}

If the latent space of a variational autoencoder (VAE) is expected to
encode physical information on the structure of a jet, then the choice
of latent space is important.  A comparison of Gaussian, Gaussian
Mixture, and Dirichlet latent spaces for detecting anomalous top jets
in a QCD sample and vice versa shows that the Dirichlet VAE (DVAE)
achieves the best performance and provides an intuitive physical
interpretation of how the jets are organised in latent
space~\cite{Dillon:2021nxw}.  The Dirichlet distribution is a family
of continuous multivariate probability distributions with the
probability density
\begin{align}
  \mathcal{D}_\alpha\left(r\right) = \frac{ \Gamma\left(\sum_i\alpha_i\right) }{ \prod_i\Gamma(\alpha_i) } \prod_i r_i^{\alpha_i-1}
  \qqquad (i=1~...~R) \; ,
\end{align}
where $R$ is the number of latent space dimensions. It is a compact
and potentially multi-modal distribution, constrained such that
$\sum_i r_i=1$, and is defined by $R$ hyper-parameters $\alpha_i>0$.
The hyper-parameters allow us to build hierarchies into the latent
space, defined by $\langle r_i\rangle=\alpha_i/\sum_i\alpha_i$.  With
$\alpha_i<1$ the distribution has $R$ modes with peaks at each of the
latent-space points $r_i=1$.  This enables the DVAE to separate the
jets into different, potentially hierarchical, modes based on their
kinematics. The DVAE can be thought of as a more powerful
deep-learning edition of more traditional topic models, such as LDA,
which have also been applied to anomaly detection problems in
high-energy physics \cite{Dillon:2019cqt,Dillon:2020quc}.

In a forward-pass through the network we need to be able to sample
from the Dirichlet distribution and calculate the KL-divergence
between two Dirichlet distributions.  This is very difficult with the
exact form of the distribution.  Therefore, we use a softmax
approximation to the Dirichlet
distribution~\cite{srivastava2017autoencoding}
\begin{align}
  r_i\sim \text{softmax}\hspace{1mm}\mathcal{N}\left(z;\tilde{\mu},\tilde{\sigma}\right)
  \qquad \text{with} \qquad
  \tilde{\mu}_i &= \log\alpha_i - \frac{1}{R} \sum_j\log\alpha_j  \notag \\
  \qquad \text{and} \qquad \tilde{\sigma}_i &=\frac{1}{\alpha_i} \left(1-\frac{2}{R}\right)+\frac{1}{R^2}\sum_j\frac{1}{\alpha_j} \; .
\end{align}
The DVAE loss function is the sum of the reconstruction loss and the
KL-divergence between the prior and the latent distribution for each
jet,
\begin{align}
\loss  = -\langle \log p_{\theta}(x|r)\rangle_{q_{\phi}(r|x)} + \beta_{\text{KL}}D_{\text{KL}}\left( q_{\phi}(r|x),\mathcal{D}_{\alpha}(r) \right) \;,
\end{align}
with a learnable encoder $q_{\phi}(r|x)$ and decoder
$p_{\theta}(x|r)$, where $\phi$ and $\theta$ are the respective
parameters.  The reconstruction loss is computed as the KL-divergence
between the input and reconstructed jet images, and the KL-divergence
between the prior and the latent-representation of the jet becomes
\begin{align}
D_{\text{KL}}\left( q_{\phi}(r|x),\mathcal{D}_{\alpha}(r) \right) = \frac{1}{2}\sum_{i=1}^R\left( \frac{\sigma_i^2}{\tilde{\sigma}_i^2} + \frac{(\tilde{\mu}_i - \mu_i)^2}{\tilde{ \sigma}_i^2} - 1-  \log\frac{\sigma_i^2}{\tilde{\sigma}_i^2} \right) \; .
\end{align}
Here, $\mu_i$ and $\sigma_i$ are the encoded means and variances in
the softmax-Dirichlet approximation for each jet.

The DVAE architecture used here is identical to the one in
Ref.~\cite{Dillon:2021nxw}.  The encoder is a neural network with 1600
inputs, a flattened $40\times40$ image, with $2R$ outputs with linear
activations, and a single hidden layer of 100 nodes with SeLU~\cite{klambauer2017self}
activations.  These outputs are the means and variances used to sample
from the softmax-Dirichlet distribution and to calculate the
KL-divergence with the prior.  The $R$-dimensional vector sampled from
the softmax-Dirichlet distribution is then passed to a decoder network
which has a very simple architecture; a 1600-dimensional output with
no hidden layers and no biases. A softmax activation is applied to the
output layer.

\subsection{Anomaly scores}
\label{sec:dvae_scores}

The Dirichlet latent space allows the VAE to separate the jets in
latent space, based on their phase-space features.  Because of the
mixture model interpretation of the DVAE and simple decoder
architecture, we can also visualise the features that the network
associates with each mixture.  From Ref.~\cite{Dillon:2021nxw} we know
that when the DVAE is trained on equal parts top and QCD jets, the
mixtures are associated with one-prong QCD-like and three-prong
top-like jets.  In contrast, for datasets with predominantly QCD jets
the learned features are one-prong and two-prong jets.

We can choose between three anomaly scores for the
DVAE~\cite{Dillon:2021nxw}; the reconstruction error, the
KL-divergence, or a latent coordinate $r_i$.  The coordinate $r_i$ is
only unambiguous for $R=2$, since $r_1=1-r_0$, while for $R>2$ there
is more than one option for the direction.  In this work we use $R=2$
with $\alpha=(1.0,0.1)$, unless otherwise specified.  We train the DVAE solely on background jets and use the
reconstruction loss as the anomaly metric, which we expect to be correlated with the density of the jets in physics space. 
We use the Adam
optimizer~\cite{Kingma:2014vow} with a learning rate of 0.01 and decay rates 
$\beta_{1,2}=(0.9, 0.99)$.  The model is trained for 300 epochs,
which is sufficient for the loss to converge. We also choose $\beta_\text{KL}=0.1$, so that
the prior has a large impact on the training.  The exact details of the implementation
and training are laid out in Tab.~\ref{tab:dvae_train}, and match
those described in Ref.~\cite{Dillon:2021nxw}.

\begin{table}[b!]
\centering
\begin{tabular}{lr}
	\toprule
	Parameter & Value \\
	\hline
	training data set size  & 100k \\
	number of epochs & 300 \\
	batch size & 2048 \\
	initial learning rate & $10^{-2}$ \\
	$\beta_{\text{KL}}$ & 0.1 \\
	$\alpha$ & (1.0,0.1) \\
	optimizer & Adam \\
	$\beta_{1,2}$ & $(0.9, 0.99)$ \\
	\bottomrule
\end{tabular}
\caption{Network and training parameters for the DVAE.}
\label{tab:dvae_train}
\end{table}

\subsection{Jet image performance}
\label{sec:dvae_img}

\begin{figure}[t]
  \includegraphics[width=0.49\textwidth]{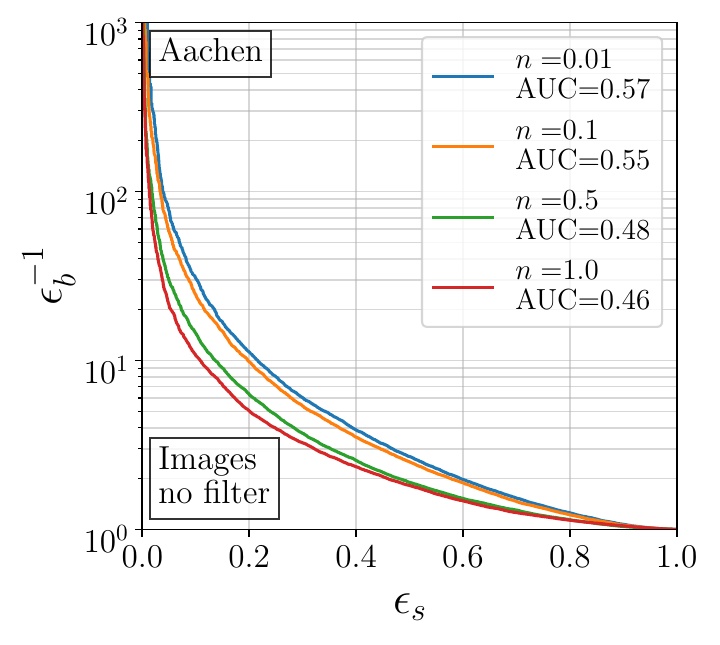}	
  \includegraphics[width=0.49\textwidth]{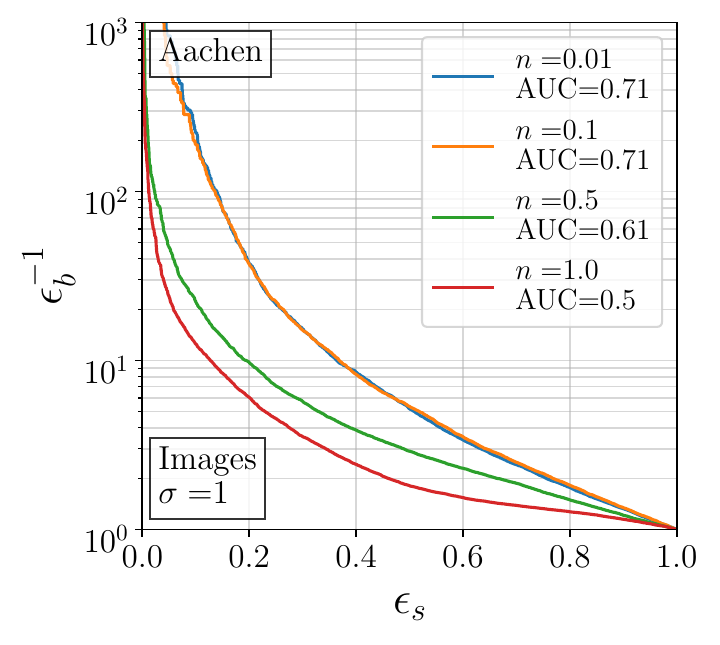}\\
  \includegraphics[width=0.49\textwidth]{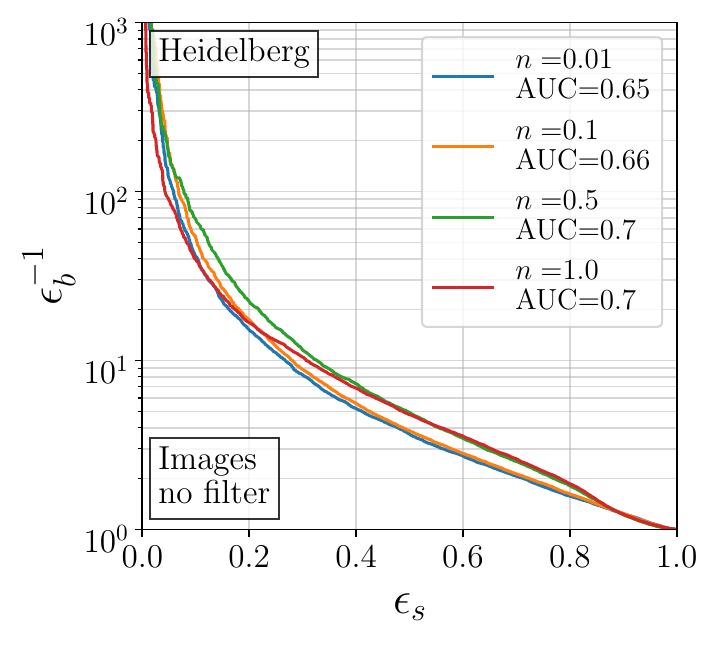}	
  \includegraphics[width=0.49\textwidth]{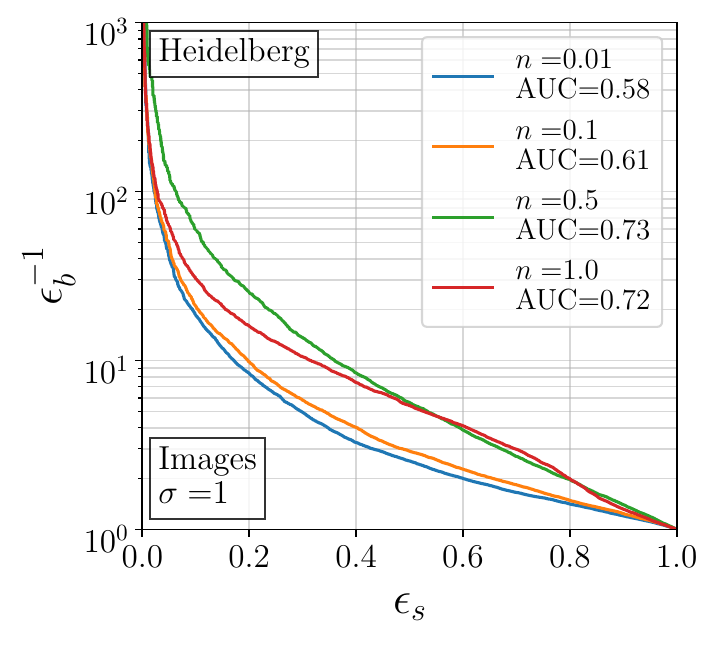}	
  \caption{ROC curves for DVAE anomaly detection using jet images on the Aachen and
    Heidelberg datasets. The reweighting power $n$ is defined in
    Eq.\eqref{eq:scalings}.}
  \label{fig:dvaeroc1}
\end{figure}

We find that the preprocessing of the jet images has a large effect on
the anomaly detection performance, with different preprocessing
parameters being optimal for different signals.  For the Aachen
dataset we find that $p_T$-reweighting,  $p_T \; \rightarrow \; p_T^n$, with $n\lesssim0.1$ and a Gaussian filter with width $\sigma\simeq1.0$ work best, resulting in an AUC~$\simeq0.71$ and a
$\epsilon_b^{-1}(\epsilon_s=0.2)\simeq37$.  While for the Heidelberg
dataset we find that $n\simeq0.4-0.6$ and $\sigma\simeq1.0$ tends to
work best, with an AUC of $\simeq0.73$ and a background suppression of
$\epsilon_b^{-1}(\epsilon_s=0.2)\simeq27$.  It is certainly not ideal
that different anomalies are best detected with different
preprocessings, although this is a recurring theme of this paper and
we argue in Sec.~\ref{sec:data} that this can be understood.  We also
note that these results do not require a fine-tuning of the
preprocessing parameters, as they are largely insensitive to order-one
changes.

The ROC curves for the DVAE are summarized in
Fig.~\ref{fig:dvaeroc1}. We see that the Gaussian filter preprocessing
has a large effect, especially for the Aachen dataset, where the AUC
drops below 0.5 in some cases without the Gaussian filter.  This is
because the anomalous features in the Aachen dataset are very sparse
and at low $p_T$, so the DVAE can and will ignore these unless they
are explicitly emphasized in the input.  We also studied a
higher-dimensional latent space, $R=3$, with $\alpha=(1.0,0.25,0.1)$,
and found no difference in the performance.  In~\cite{Dillon:2021nxw}
it was found that for the top tagging the mixture weights or latent
coordinates $r_i$ can provide good performance in anomaly detection
even when the anomalous jets are less complex than the background.
Here we find a similar behaviour, but the performance is not as good
as for the reconstruction error.

\subsection{EFP performance}
\label{sec:dvae_efp}

\begin{figure}[t]
  \includegraphics[width=0.49\textwidth]{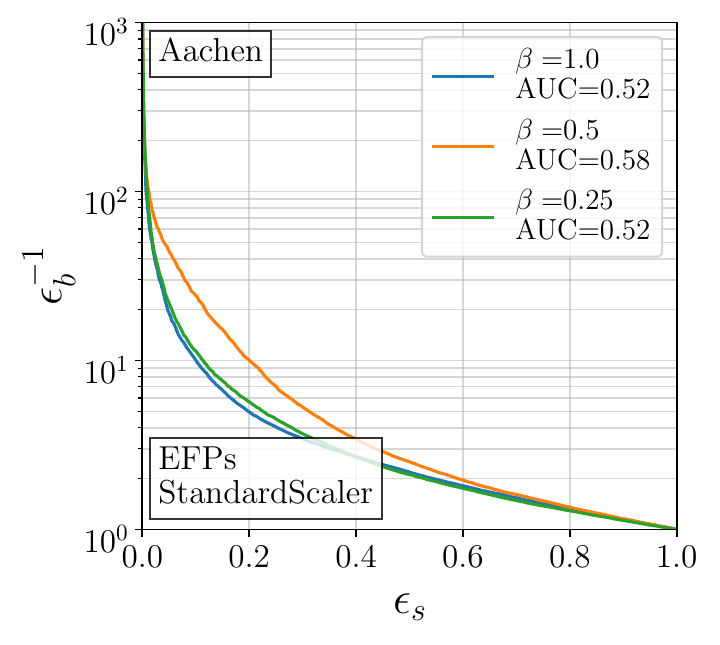}	
  \includegraphics[width=0.49\textwidth]{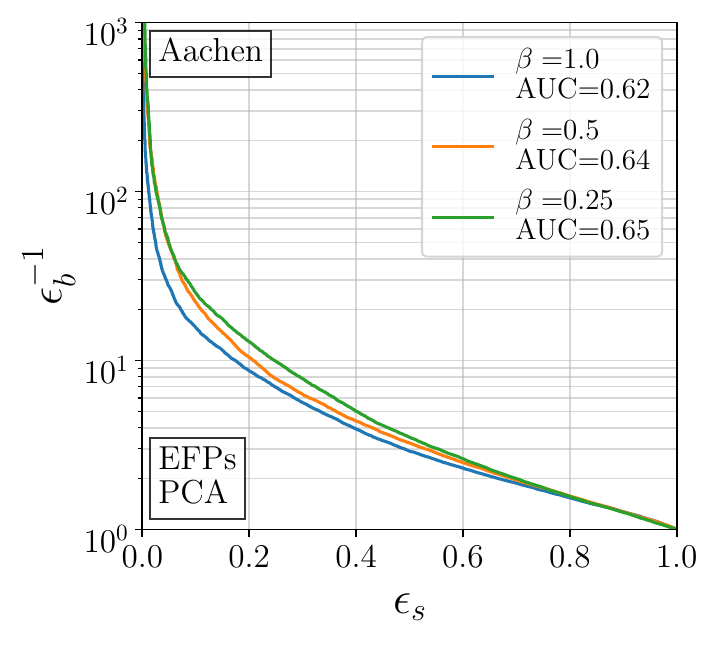}\\
  \includegraphics[width=0.49\textwidth]{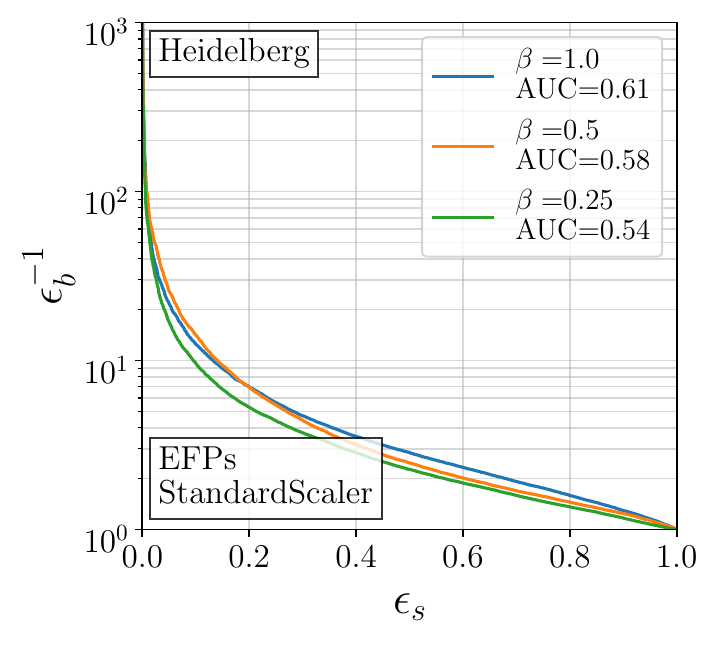}	
  \includegraphics[width=0.49\textwidth]{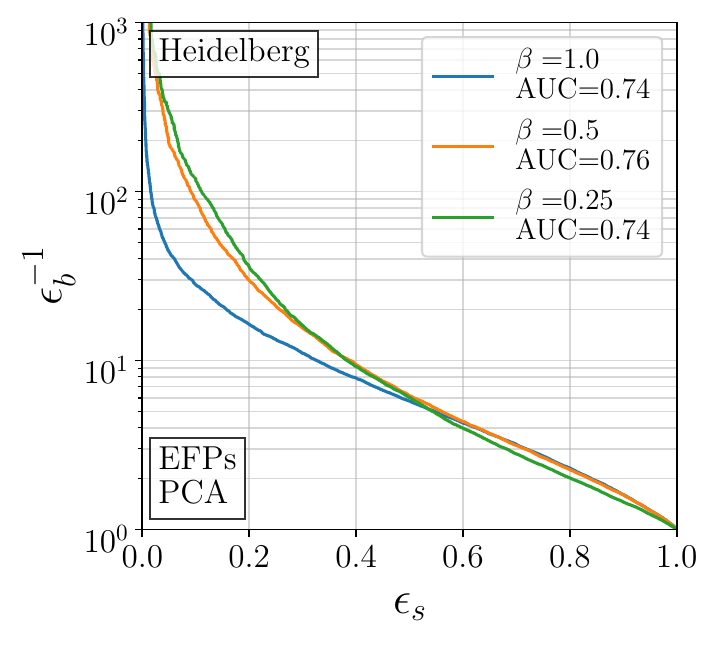}	
  \caption{ROC curves for DVAE anomaly detection using EFPs on the Aachen and
    Heidelberg datasets ($\kappa=1$).}
  \label{fig:dvaeroc2}
\end{figure}

The DVAE aims to extract physical information from the images, but it
is of course possible to use EFPs as input rather than images.  This
way it might be easier for the algorithm to characterize the
background and better identify anomalies.

As discussed in Sec.~\ref{sec:data_pre} we consider two preprocessings
for EFPs: (i) standard scaling where we set the mean to zero and the
standard deviation to unity, not removing the strong correlation
between EFPs, and (ii) using PCA components, still with zero mean, but
an identity covariance matrix between features.  These are performed
with the StandardScaler and PCA routines in
scikit-learn~\cite{scikit-learn}.  We have also learned from the
previous section that reweighting physical inputs heavily influences
the anomaly detection performance. Here we investigate what happens
when we train our DVAE on the first eight EFPs and we preprocess them
using $z_i \to z_i^\kappa$ and $R_{ij} \to R_{ij}^\beta$, see
Eq.\eqref{eq:def_kappabeta}, in analogy to the pixel reweighting used
before.  We explore a range of values for $\beta$ and $\kappa$, as
well as having either two or three mixtures in the latent space,
i.e. $R=2$ or $R=3$.  In agreement with the DVAE trained on images, we
find that the results with $R=2$ and $R=3$ are essentially the same.
A summary of the DVAE performance for $\kappa=1$ and
$\beta=0.25~...~1$ are shown in Fig.~\ref{fig:dvaeroc2}. First, we see
that standard scaling without decorrelating the EFPs essentially
fails. Once we include the decorrelation step, we find that for both
the Aachen and Heidelberg datasets using EFPs with $\kappa=1$ and
$\beta=0.25~...~0.5$ gives the best anomaly detection. This is in
complete agreement with our observation from k-means clustering and
the image-based DVAE. Directly comparing the performance of the
image-based and EFP-based DVAEs we find that in our current setup the
images work slightly better for the Aachen dataset, while the EFPs are
more efficient in extracting the hard substructure of the Heidelberg
dataset.

\section{INN}
\label{sec:inn}

A density estimation based on neural networks can be obtained using
normalizing flows~\cite{dinh2017density, kingma2018glow}, specifically
their invertible neural network (INN)
incarnation~\cite{inn,cinn}. INNs are neural networks which learn
bijective mappings between a physics and a latent space completely
symmetrically in both directions. They allow access to the Jacobian
and both directions of the mapping, linking density estimation in the
physics and latent spaces in a completely controlled manner.  We have
used the flexible INN setup successfully for precision event
generation~\cite{Bellagente:2021yyh,Butter:2021csz}, unfolding
detector effects~\cite{Bellagente:2020piv}, and QCD or astro-particle
inference~\cite{Bieringer:2020tnw,Bister:2021arb}.

\begin{table}[b!]
\centering
\begin{tabular}{lr}
	\toprule
	Parameter & Value \\
	\hline
	training data set size  & 100k \\
	number of epochs & 300 \\
	batch size & 512 \\
	initial learning rate & $10^{-4}$ \\
	learning rate decay & 0.98 \\
	initial noise width & 0.5 \\
	noise decay & 0.95 \\
	optimizer & Adam \\
	$\beta_{1,2}$ & $(0.9, 0.999)$ \\
	scaling & PCA \\
	\bottomrule
\end{tabular}
\caption{Network and training parameters for the INN.}
\label{tab:inn_train}
\end{table}

For this straightforward application we use simple, affine coupling
blocks~\cite{dinh2017density} combined with random, but fixed,
orthogonal transformations.  In an affine coupling block, the input
dimensions are split into two halves, $x_{1,2}$. The first half is
passed through a subnet which learns two functions $s(x_1)$ and
$t(x_1)$. The second half is transformed by an element-wise
multiplication $(\odot)$ and an element-wise addition,
\begin{align}
\begin{pmatrix} z_1 \\ z_2 \end{pmatrix} = \begin{pmatrix} x_1 \\ x_2\odot e^{\text{s}(x_1)} + \text{t}(x_1) \end{pmatrix}
\qquad \Leftrightarrow \qquad
\begin{pmatrix} x_1 \\ x_2 \end{pmatrix} = \begin{pmatrix} z_1 \\ (z_2 - \text{t}(z_1)) \odot e^{-\text{s}(z_1)} \end{pmatrix} \; .
\label{eq:affine}
\end{align}
The Jacobian of this mapping is
\begin{align}
  J = \begin{pmatrix} \one & 0 \\ \partial_{x_1}z_2  & \text{diag} \; e^{\text{s}(x_1)} \end{pmatrix}
  \qquad \Rightarrow \qquad \log |J| =\sum s(x_1) \; .
  \label{eq:jacobian}
\end{align}
We use soft clamping to avoid instabilities in the
training~\cite{inn}, replacing $s(x_1) \to 2 \tanh s(x_1)$ in
Eqs.\eqref{eq:affine} and~\eqref{eq:jacobian}. The random
transformations ensure that in each coupling block the information is
split differently. They are easily invertible, and their Jacobian is
one.

If $f$ is the mapping between physics space and the INN latent space,
and $q$ is the prior distribution in latent space, then the learned
distribution in physics space $p$ can be written as $p(x) = q(f(x)) \;
\abs{J(x)}$.  The loss function should become minimal if the learned
distribution in physics space $p$ matches the true distribution
$p_\text{true}$.  Therefore, we would like to minimize the
KL-divergence between $p$ and $p_\text{true}$,
\begin{align}
	D_\text{KL}(p,p_\text{true}) =
	\int dx \; p_\text{true}(x) \; \log \frac{p_\text{true}(x)}{p(x)} \;.
\end{align}
Since we know $p_\text{true}$ only from samples $\{x_i\}$, this is
difficult to achieve. 
Moreover, we can split the KL-divergence into
the self-entropy of $p_\text{true}$, which does not depend on $f$, and
the negative log-likelihood. The loss function is this negative
log-likelihood,
\begin{align}
	\loss
	&= -\int  dx \; p_\text{true}(x) \; \log p(x)
	= -\int dx \; p_\text{true}(x) \;
	\Big[ \log q(f(x)) + \log |J(x)| \Big] \notag \\
	&\approx - \frac{1}{N}\sum_{i=1}^{N} \Big[ \log q( f(x_i)) + \log \abs{J(x_i)} \Big] \; ,
	\label{eq:inn_loss}
\end{align}
which we can evaluate without having an explicit form of
$p_\text{true}$. This loss function automatically ensures that the
latent distribution follows the prior~\cite{inn}.

Learning a density from jet images faces the challenge that the active
pixels are distributed very sparsely, which means that most of the
large number of physics space dimensions do not carry any
information. This inflates the number of dimensions in the image-pixel
space, while we know that the relevant number of dimensions describing
the jets is much smaller. For the bijective INN we are interested in
limiting the number of physical and latent dimensions, so we simplify
our task by using the eight EFPs up to order $d = 3$ introduced in
Sec.~\ref{sec:data}. The 8-dimensional physics space is then mapped on
an 8-dimensional Gaussian latent space. Our architecture consists of
24 affine coupling blocks, each followed by a random orthogonal
transformation. Our subnets are each a fully connected network with
one hidden layer of 512 nodes and ReLu activation. The output layer
has no activation. The network parameters are summarized in
Tab.~\ref{tab:inn_train}.

We train our network for 300 epochs using the Adam
optimizer~\cite{Kingma:2014vow} and an initial learning rate of
$10^{-4}$. The learning rate is then reduced every epoch by 2\%.  To
stabilize the training, we apply a PCA as described in
Sec.~\ref{sec:data_pre}. Since the PCA is also an invertible
transformation with a computable Jacobian, we can still evaluate the
density in the original EFP space. Also we help the training by adding
Gaussian noise to the training data, where we reduce the standard
deviation of the noise by 5\% every epoch, so that it had no effect by
the end of the training. Further training details can be found in
Tab.~\ref{tab:inn_train}.

\subsection{Anomaly scores}
\label{sec:inn_scores}

\begin{figure}[t]
  \includegraphics[width=0.33\textwidth]{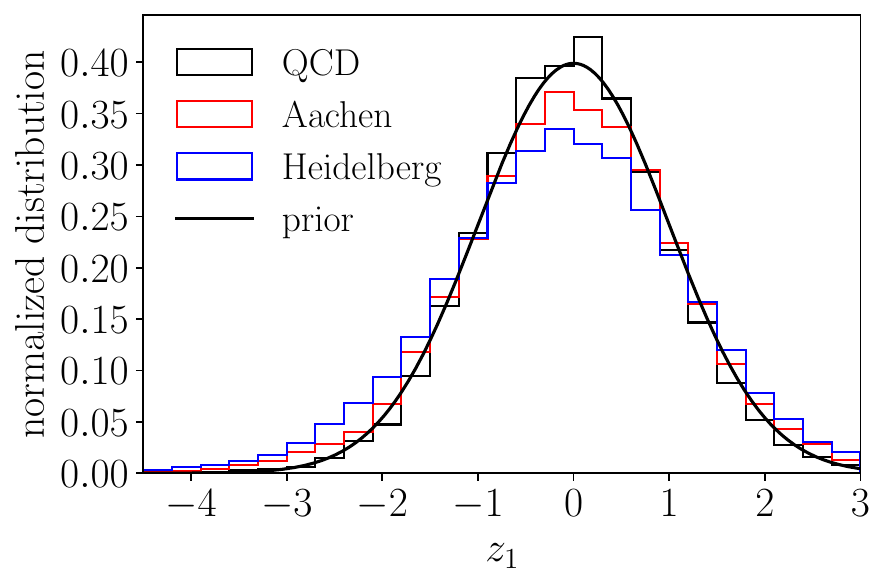}
  \includegraphics[width=0.33\textwidth]{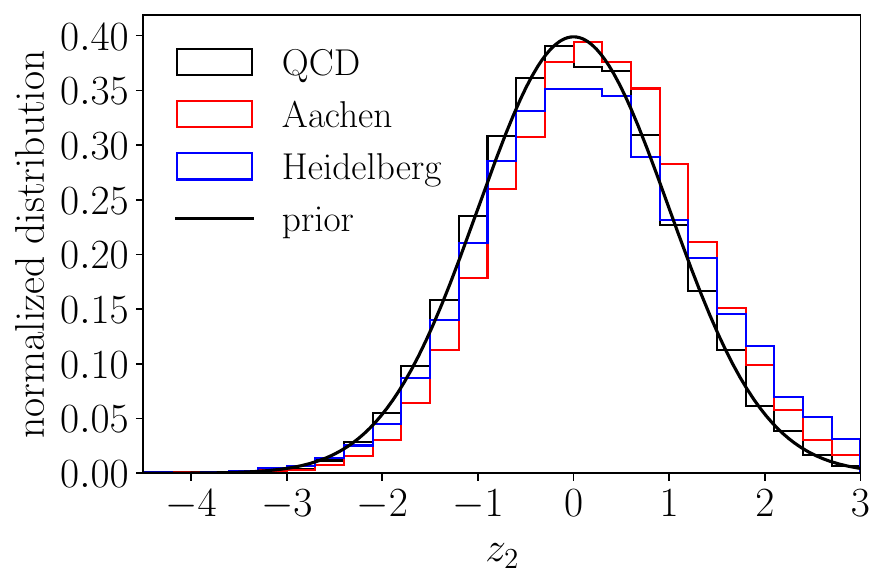}
  \includegraphics[width=0.33\textwidth]{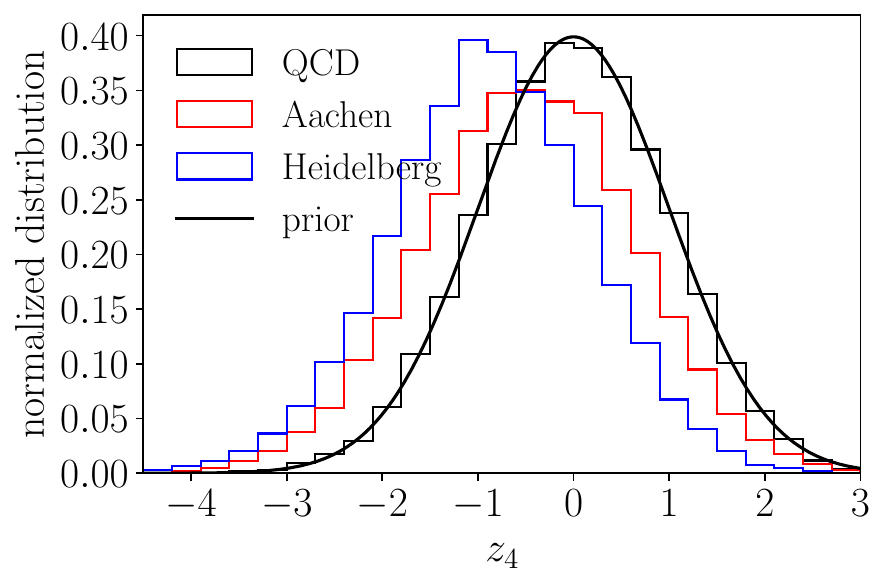} \\
  \includegraphics[width=0.33\textwidth]{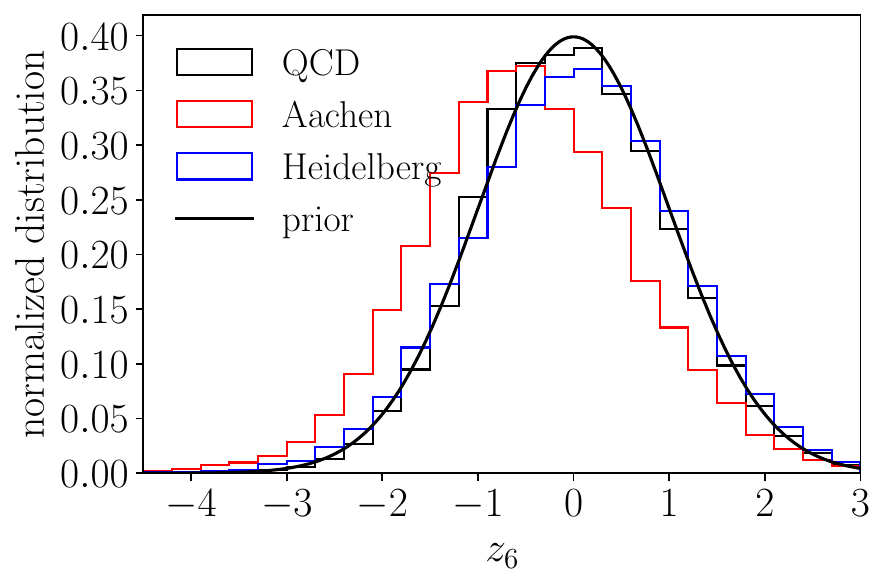}
  \includegraphics[width=0.33\textwidth]{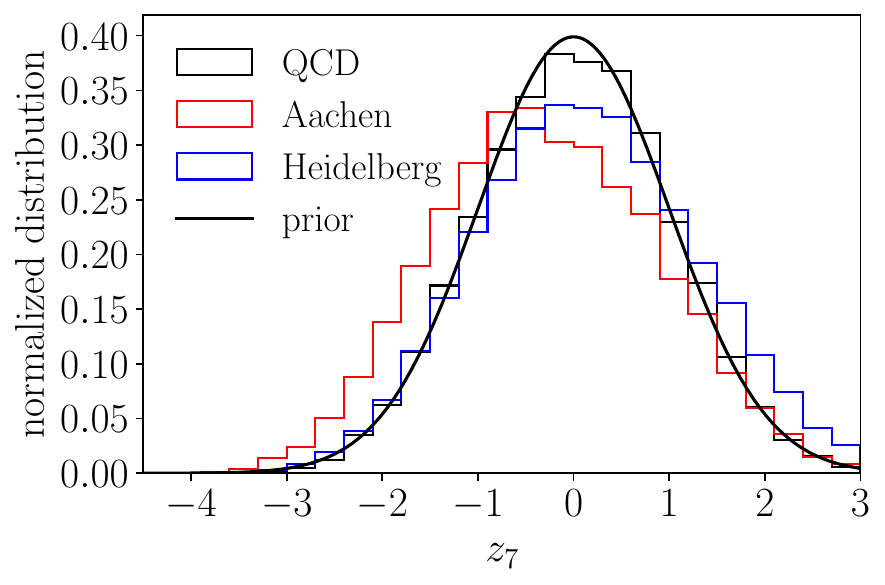}
  \includegraphics[width=0.33\textwidth]{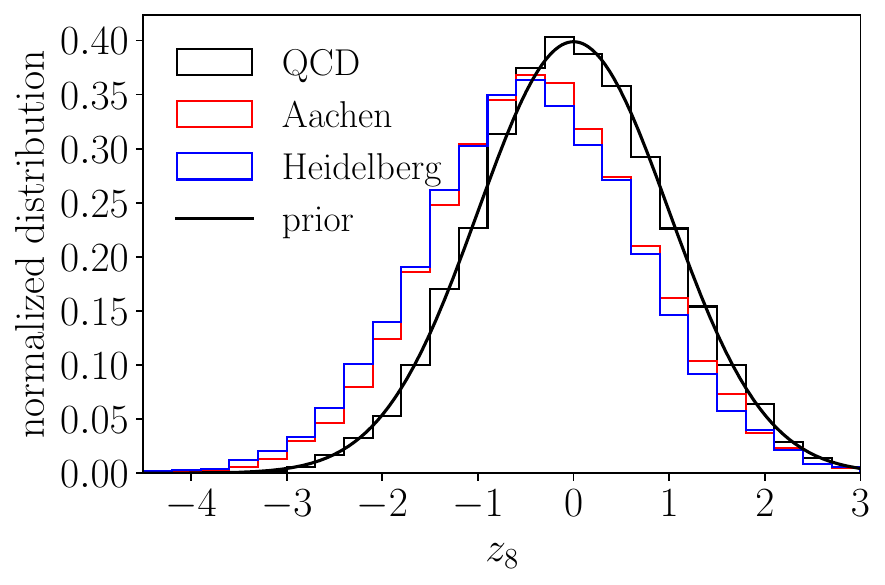}
  \caption{Example distribution of background and signal jets in the
    INN latent space, after training on background only.}
  \label{fig:inn_latent}
\end{figure}

The INN allows us to estimate the density associated to where a
particular jet lies in the space of physical observables we use to
characterize it.  It does this by constructing a Gaussian latent space
for the jets along with a learned Jacobian to correctly account for
the density in the transformation.  The idea behind using an INN to
search for anomalous jets is that we can look for jets located in low
density regions of either physics space or the Gaussian latent space,
with the Jacobian connecting the two. 

Using the density of a given jet in physics space as the anomaly
score, based on information from both the latent Gaussian space and
the Jacobian is well motivated and has a clear definition. It can also
be interpreted as a standard observable, defined in physics space, but
numerically represented through a neural network benefiting from a
specific latent space.

In the latent space, we expect the INN to ensure that the background
forms a Gaussian distribution, including the exponentially suppressed
tails away from the mean, or typical jet patterns.  
Thus we can use the distance to the center as anomaly score,
which is monotonously related to the likelihood for a
multi-dimensional Gaussian. In Fig.~\ref{fig:inn_latent} we show the
latent space distribution for the QCD training dataset and the Aachen
and Heidelberg dark jets. While the QCD distributions follow the
Gaussian prior closely, as expected when training on QCD jets only,
both signal datasets differ from the prior in some of the latent
directions.  The drawback of this method is that the training does not
guarantee the signal to end up in the tales of the latent distribution. 

\begin{figure}[t]
  \includegraphics[width=0.49\textwidth]{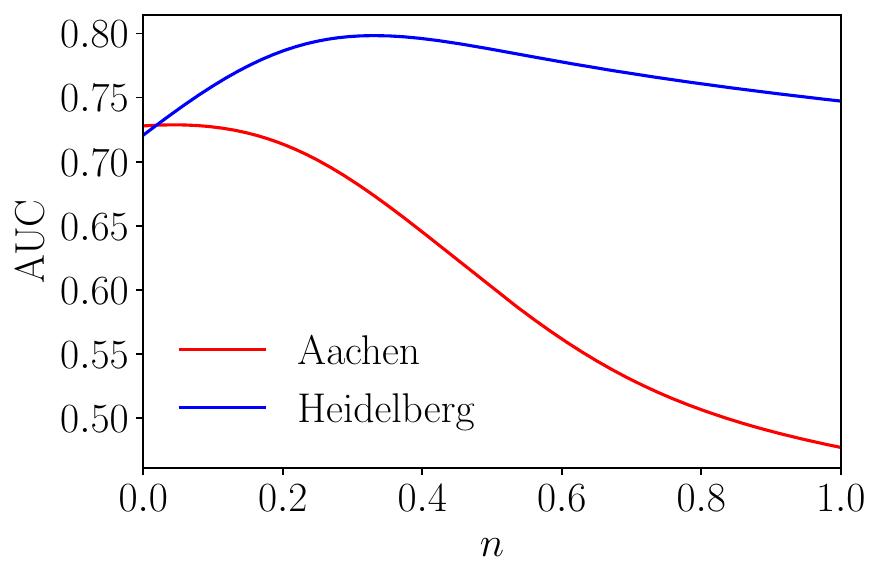}
  \includegraphics[width=0.49\textwidth]{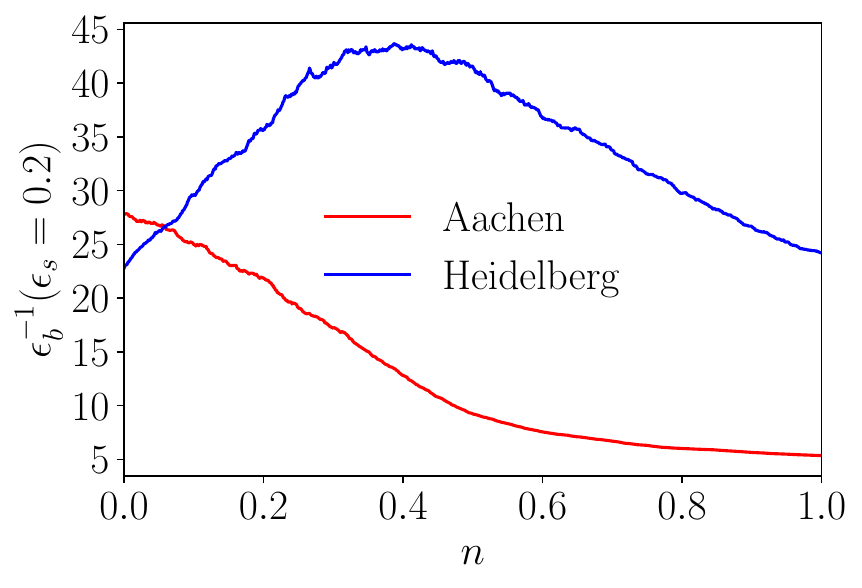}
  \caption{Effect of the input scaling $x\to x^n$ on the AUC and the
    inverse mistag rate at 20\% signal efficiency.}
  \label{fig:inn_remap}
\end{figure}

\subsection{Performance}
\label{sec:inn_results}

Applying the INN to EFPs with $\kappa=\beta=1.0$ and using the
negative log-likelihood in physics space as an anomaly score leads to
a strong bias to detect more complex jets as anomalous. This is due to
the fact that jet density is particularly high at low EFP values and
jets with less structure have lower EFP values. This bias can be
compensated for by using a reweighting that leads to more uniform
distributions. We study an element-wise exponentiation $g(x)=x^n$
applied to the EFP before the PCA, so the distribution in the
$x'$-space is $p'(g(x))=p(x)\cdot\abs{J_g(x)}^{-1}$. The Jacobian of
this reweighting is again diagonal, and the negative log-likelihood of
a single sample transforms like
\begin{align}
  \loss \rightarrow \loss
  + \sum \log \left|\frac{d g}{dx} \right|
  = \loss + (n-1) \sum \log x + \sum \log n  \;,
\end{align}
where the sums go over the eight input dimensions. We drop the last term since it is independent of $x$ and will therefore have no
effect on tagging. Then for $n=0$ this transformation corresponds to the
reweighting $g(x)=\log(x)$.  This technique for reweighting would be
analogous to training the DVAE, or any autoencoder, on reweighted
inputs, and then applying the reweighting to the input and
reconstructed images when evaluating the anomaly score.

\begin{figure}[t]
  \includegraphics[width=0.49\textwidth]{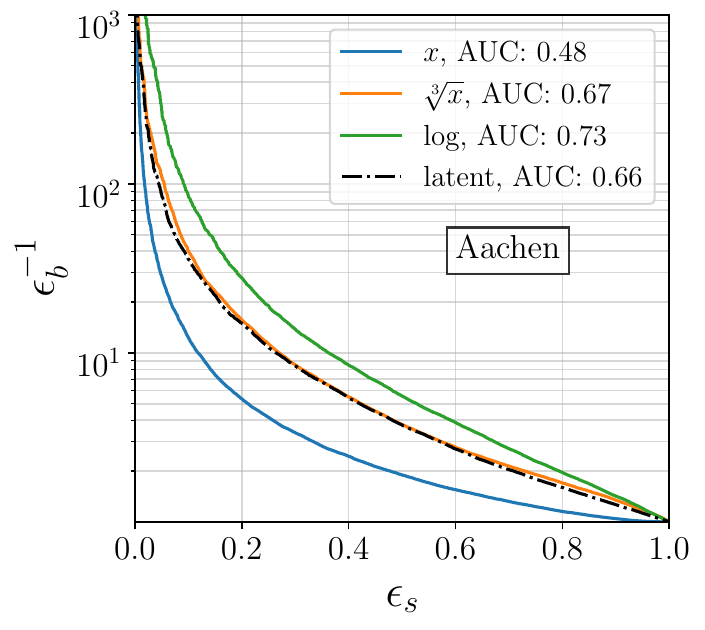}
  \includegraphics[width=0.49\textwidth]{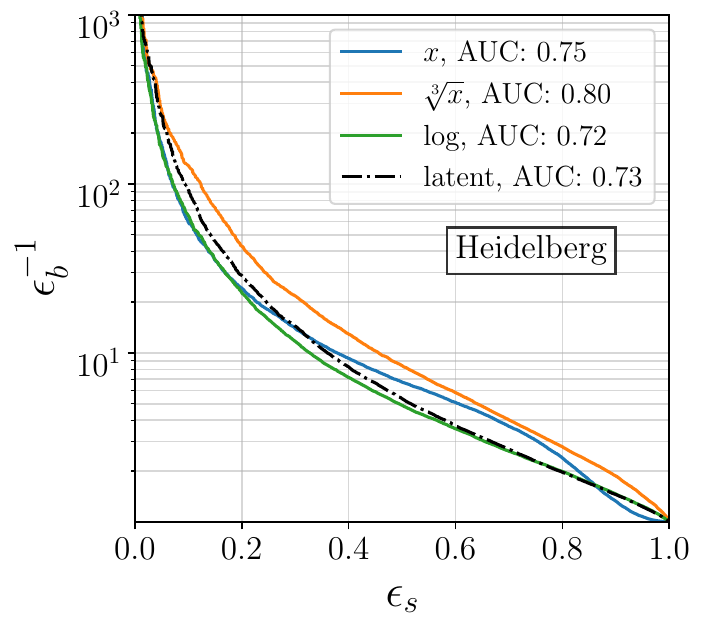}
  \caption{ROC curves for INN anomaly detection on the Aachen and
  	Heidelberg datasets, using density based tagging in physics and latent
  	space. The density in physics space is effected by reweighting while the
  	density in latent space is not.}
  \label{fig:results_inn}
\end{figure}

Fig.~\ref{fig:inn_remap} shows the effect of this reweighting for $n
\in [0,1]$ on the AUC and the inverse mistag rate at 20\% signal
efficiency. The point $n=1$ means no reweighting at all. It can be
seen that a signal with less structure than the background like the
Aachen data set needs low values of $n$ to be detected, as
expected. Fig.~\ref{fig:results_inn} shows the ROC curves for
density based tagging in physics and in latent space. Only the density
in physics space is affected by the reweighting. For the plots we
choose three representative values for $n$.  Despite the INN not being
retrained on EFPs with different reweightings, the anomaly detection
performance drastically improves for different $n$.  This implies that
the INN is learning the low-density regions in physics space well, and
that difficulties in detecting the anomalies are related to the choice
of observables rather than the network training. A remaining challenge
for the INN-based anomaly search is the choice of the reweighting, as
the log-reweighting works best for the semi-visible Aachen dataset,
while the cubic-root-reweighting gives the best results for the
Heidelberg dataset.

\section{Outlook}
\label{sec:outlook}

Searches for anomalies, or hints for physics beyond the Standard
Model, are a large part
of the LHC program and essentially all
particle physics experiments. We have studied how this analysis goal
can be pursued with modern machine learning tools and concepts,
specifically unsupervised learning. New ways to disentangle potential
signals from background can be applied at the LHC in many ways, at the
trigger and analysis stages, on jets or other specific analysis
objects and on whole events, with training on data and on simulation. 

In this paper we have studied jets, which are an excellent path to
understanding modern anomaly search tools, because they are
theoretically well understood, can be simulated in large numbers, and
come with huge established data samples.  We do know that classic
out-of-distribution searches will not work in LHC
physics, because in high-statistics jet or event samples the
backgrounds eventually populates all possible phase space
configurations. This suggests to consider anomaly searches in relation
to probability density estimation. This task is reflected in our
choice of two benchmark signals, both inspired by a dark or invisible
sector interfering with QCD showering. The Aachen dataset features
semi-visible jets, where part of the shower products are not visible,
while the Heidelberg dataset features massive decays inside a fully
visible jet.\medskip

We studied three different ways of searching for such dark jets in a
QCD background sample, assuming that we can train in an unsupervised
way on background only or with negligible signal contamination:
\begin{itemize}
  \item[$\cdot$] K-means is a classic ML algorithm, which we use to
    define a set of different anomaly scores. It works on jet images
    and allows us to study density estimation in the corresponding
    high-dimensional spaces. For a well-motivated choice of
    hyperparameters the performance is stable and competitive with
    more modern deep learning methods.  Anomaly searches with k-means
    clustering benefit from a reweighting of the image pixels as part
    of the preprocessing, as do all methods we have analysed.
  \item[$\cdot$] Dirichlet VAEs are a modern ML-algorithm which
    constructs a multi-modal latent space, allowing us to define
    anomaly scores in the physics space or in the latent space. The reconstruction loss is
    used as the anomaly score, which is expected to be correlated with 
    the density of the jets in physics space.
    They work on
    jet images and on lower-dimensional representations like energy
    flow polynomials (EFPs). We found that Dirichlet VAEs also require
    preprocessing, and while 
    it is possible to use a latent-space anomaly score,
    the reconstruction
    loss performs better for our benchmark signals.
  \item[$\cdot$] Normalizing flows, specifically INNs, construct a
    bijective and traceable map between the physics and latent
    spaces. This means they are ideally suited for density estimation
    but they work best on lower-dimensional data representations, like
    decorrelated EFPs. INNs have the advantage that we have full
    control over the link between the two spaces, and we found the
    density estimate in physics space to provide the best-performing
    anomaly score.
\end{itemize}
\medskip

Our three very different algorithms have very different requirements, strengths, and
caveats. Once understood, they all show similar performance on both
datasets. Preprocessing is important for all of them, and has a very
significant impact on the performances. 
This is because we define the anomalies as those jets that lie in low 
density regions of physics space, and the preprocessing alters this density,
therefore it changes how the anomalies are defined.
We finally note that our
Aachen and Heidelberg datasets are challenging datasets for anomaly
detection, compared for example to top-tagging, which has served as a
benchmark for assessing the performance ML methods. The datasets will be 
available as public benchmarks for existing and improved anomaly
detection algorithms.

\begin{center} \textbf{Acknowledgments} \end{center}

\noindent This research is supported by the Deutsche
Forschungsgemeinschaft (DFG, German Research Foundation) under grant
396021762 -- TRR~257: \textsl{Particle Physics Phenomenology after the
  Higgs Discovery}, grant 400140256 - GRK 2497: \textsl{The physics of
  the heaviest particles at the Large Hardon Collider} and through
Germany's Excellence Strategy EXC 2181/1 - 390900948 (the Heidelberg
STRUCTURES Excellence Cluster).

\begin{center} \textbf{Notes} \end{center}
%
The codes for the k-means clustering, the DVAE and the INN can be found at
\url{https://github.com/IvanOleksiyuk/jet-k-means},
\url{https://github.com/bmdillon/jet-mixture-vae} and
\url{https://github.com/ThorstenBuss/jet-inn},
respectively.

\bibliographystyle{SciPost-bibstyle-arxiv}
\bibliography{literature2}
\end{document}